\documentclass[lettersize,journal]{IEEEtran}
\usepackage{amsmath,amsfonts}
\usepackage{algorithmic}
\usepackage{comment}
\usepackage{algorithm}
\usepackage{array}
\usepackage[caption=false,font=scriptsize,labelfont=sf,textfont=sf]{subfig}
\usepackage{textcomp}
\usepackage{stfloats}
\usepackage{url}
\usepackage{verbatim}
\usepackage{graphicx}
\usepackage{cite}
\usepackage{bm} % 加载 bm 包
\usepackage{float}
\usepackage{amsmath} % 加载 amsmath 包
\usepackage{mathrsfs}
\usepackage{mathalpha}
\usepackage{booktabs}  % 专业表格线
\usepackage{multirow}  % 多行合并
\usepackage{array}     % 表格格式控制
\usepackage{pifont} %圈字符
\usepackage{tcolorbox} %颜色包

\usepackage[numbers,sort&compress]{natbib} % 使用 natbib 包

\usepackage{indentfirst} % 自动处理首段缩进
\usepackage{titlesec}

\titlespacing*{\section}{0pt}{6pt}{4pt}
\titlespacing*{\subsection}{0pt}{5pt}{3pt}  % 调整 subsection 间距
\titlespacing*{\subsubsection}{0pt}{4pt}{2pt}  % 调整 subsection 间距

\DeclareMathOperator{\sigmoid}{sigmoid}
\newcommand{\sneq}{\mathrel{\mkern-5mu}=\mkern-5mu} % 在导言区定义

\begin{document}
	
	%SoM-MTM: Synesthesia of Machines (SoM)-Driven Mask Token Model for Cooperative Perception Enhancement over Packet Loss Channels
	\title{SoM-MTM: Synesthesia of Machines (SoM)-Driven Masked Token Model for Cooperative Perception over Packet Loss Channel}
	\author{Haozhen Li,~\IEEEmembership{Graduate Student Member,~IEEE,}
	Rongqing Zhang,~\IEEEmembership{Senior Member,~IEEE and} \\
	Xiang Cheng,~\IEEEmembership{Fellow,~IEEE}
		% <-this % stops a space
		\thanks{Manuscript received xxxx. (\textit{Corresponding author: Xiang Cheng.})}  %致谢信息待定
		% <-this % stops a space
		\thanks{Haozhen Li and Xiang Cheng are with the State Key Laboratory of Photonics and Communications, School of Electronics, Peking University, Beijing 100871, China (email: \{pkuimlhz, xiangcheng\}@pku.edu.cn).}  %致谢-作者信息
		\thanks{Rongqing Zhang is with the Intelligent Transportation Thrust, The Hong Kong University of Science and Technology (Guangzhou), Guangzhou 511453, China (email: rongqingz@hkust-gz.edu.cn).}  %致谢-作者信息
		%\thanks{Manuscript received April 19, 2021; revised August 16, 2021.}  %终稿相关的信息暂无
	}
	
	% The paper headers % header信息暂无
	%\markboth{Journal of \LaTeX\ Class Files,~Vol.~14, No.~8, August~2021}%
	%{Shell \MakeLowercase{\textit{et al.}}: A Sample Article Using IEEEtran.cls for IEEE Journals}
	% \IEEEpubid{0000--0000/00\$00.00~\copyright~2021 IEEE}    %IEEE bulid信息暂无
	% Remember, if you use this you must call \IEEEpubidadjcol in the second
	% column for its text to clear the IEEEpubid mark.
	
	\maketitle
	\begin{abstract}
		To support the large-scale and heterogeneous visual cooperative perception (CP) demands in next-generation mobile networks, intelligent and efficient sensory data transmission is a critical challenge.
		Under the emerging convergence of communication networks and agentic artificial intelligence (AI), existing research emphasizes utilizing end-to-end neural networks to simplify communication modules, which has shown promising potential for CP.
		However, these studies are still limited to specific channel models,  cooperation modes, and perception tasks, failing to fully leverage powerful visual processing approaches to enhance universality.
		%介绍方法
		To address this, we propose a Synesthesia of Machines (SoM)-driven Masked Token Model, referred to as SoM-MTM, as a plug-and-play paradigm for generic visual CP.
		%Inspired by masked image modeling methods such as MAE, it possesses great perceptual context learning capabilities to achieve efficient information carrying, making it highly compatible with the packet loss channel.
		Inspired by masked image modeling methods such as MAE, it possesses great perceptual context learning capabilities to recover distorted features over packet loss channels, thereby improving information carrying efficiency.
		Building upon Swin Transformer, SoM-MTM further embeds prior masked information through an External Routing MoE mechanism, maximally repairing and enhancing environmental perception features during cooperation.
		%一句话性能
		Comprehensive experimental results confirm that SoM-MTM can consistently enhance perception performances on various tasks, especially strong generalization to unseen scenarios, while maintaining favorable model cost and scalability.
	\end{abstract}
	
	\begin{IEEEkeywords}
		SoM, perceptual contextual learning, MAE, packet loss channel, plug-and-play.
	\end{IEEEkeywords}
	
	\section{Introduction}\label{Sec-1}
	
	\IEEEPARstart{N}{ext-generation} networks are expected to support intelligent service demands of massive mobile agents in typical scenarios covering the Internet of Things (IoT), smart factories, vehicle-to-everything (V2X), and autonomous driving (AD). 
	With the emergence of multi-agent and swarm intelligence \cite{Concept_MultiAgent}, cooperative perception (CP) \cite{CP_Survey} has gradually become a cornerstone of diverse intelligent functionalities. 
	Based on accurate environmental perception and sensory data transmission among agents, it can effectively undertake tasks including teleoperation \cite{Teleoperation}, collaborative simultaneous localization and mapping (SLAM) \cite{CSLAM}, target re-identification \cite{ReID}, and beyond-line-of-sight object detection \cite{V2XViT}, thereby empowering upper-layer decision-making and task execution.
	
	Current CP systems generally operate under the following paradigm: agents collect and process the sensory data, while the network enables reliable data transmission, achieving accurate environmental reconstruction or semantic understanding such as category and location information, across agents. As the number of nodes increases and system operating conditions expand in future scenarios \cite{6GforV2X}, data transmission will impose significantly greater pressure on communication networks, posing severe challenges to this paradigm.
	This is because it heavily relies on error-checking and retransmission mechanisms, where transmission failure inevitably leads to extra requests and waiting via protocols like Automatic Repeat reQuest (ARQ) \cite{ARQ}. Although certain studies like V2VNet \cite{V2VNet} and Where2comm \cite{Where2comm} can partially alleviate the transmission burden through sensory data compression, the resulting \textbf{retransmission latency and long-tail distribution become unacceptable} when network connectivity is heterogeneous and complex, or wireless channel quality is poor.
	
	%单线逻辑，说可以去除重传
	Compared with traditional communication paradigms, networks in the B5G/6G era increasingly emphasize the convergence with AI, with communications among machines or mobile agents no longer necessarily requiring exactly reliable data transmission.
	Joint Source-Channel Coding (JSCC) \cite{DeepJSCC} represents such technological shift, which can simplify communication modules through semantic communication \cite{Concept_SC} rather than guaranteeing accuracy of every bit, with deep neural networks playing a critical role in this process.
	%so as to enable efficient information carrying under limited channel conditions.
	In this context, it is highly advisable to integrate more advanced and powerful AI tools into CP over realistic communication links, aiming to improve visual information carrying efficiency across general scenarios and tasks.
	
	Motivated by masked image modeling (MIM) for self-supervised visual representation, we introduce the concept of \textbf{perceptual contextual learning into CP over packet loss channels}. 
	Specifically, by organizing each transmitted token as an individual packet, packet loss can be naturally modeled as token masking, where the successfully received tokens provide partial observations for perceptual feature recovery.
	Similar to self-supervised visual representation such as masked autoencoders (MAE) \cite{MAE} that operate on raw images, the proposed paradigm achieves remarkable contextual learning capability in the token space. 
	This makes it natively compatible with packet loss channels, reducing the dependence on retransmission mechanisms, which can greatly reduce data communication delay and jitter, improve cooperation efficiency, particularly suitable for complex and dynamic channel conditions as well as latency-sensitive perception applications.

	%引入SoM
	On this basis, we aim to organically combine visual masked learning with wireless communication, making approaches like MAE more suitable for data transmission in CP. 
	Inspired by human synesthesia, in which the stimulation of one sense organ will automatically evoke another sense organ to jointly
	perform cognitive tasks, the Synesthesia of Machines (SoM) is
	proposed in \cite{SoM_COMST}.
	It aims to extract compact and robust features through AI-native neural network models, empowering general communication and perception tasks and achieving intelligent integration \cite{SoM_TNSE}. 
	Under the guidance of SoM, our goal is to deeply couple perceptual context learning with packet loss channel condition, constructing a plug-and-play CP scheme that is sufficiently efficient and broadly applicable.
	
	Therefore, we develop a \textbf{SoM-driven Masked Token Model as a plug-in module} for cooperative perception over packet loss channels, termed \textbf{SoM-MTM}. 
	We perform extensive experiments to confirm that it outperforms existing data transmission frameworks for CP across different tasks and channel conditions. 
	In typical settings, it can attain performance improvements of 8\% in in-distribution cases and 23\% in out-of-distribution cases, while maintaining favorable scalability and cost-performance trade-off.
	Accordingly, our SoM-MTM is able to underpin more efficient and scalable CP with endogenous intelligence. % in application scenarios.
	The key contributions of our work are summarized in the following aspects:
	\begin{itemize}
		\item To tackle connection efficiency challenges among mobile multi-agents in CP, we propose a tightly-coupled architecture that bridges visual representation with data transmission under the guidance of the SoM paradigm.
		We specifically design a generic cooperation process apt for packet loss channels, which greatly reduces the dependence on retransmission mechanisms and improves cooperative perception efficiency and robustness through compact information carrying and strong contextual learning.
		\item To adapt masked image modeling to CP, we utilize packet status information as an additional prior and embed it into SoM-MTM. We incorporate it as External Routing to improve the Mixture of Experts technique (ERMoE), flexibly enhancing Swin Transformer via ERMoE-MHSA and ERMoE-FFN. Consequently, our SoM-MTM exhibits strong adaptability to limited channel conditions, particularly superior generalization to unseen scenarios.
		\item To ensure universal significance and deployment potential, we formulate SoM-MTM as a plug-and-play module, which facilitates effective knowledge integration while maintaining reasonable model overhead.
		Experiments on different tasks and configurations validate its broad applicability and scalability, enabling cost-effective migration and deployment in diverse networked applications.
	\end{itemize}

	%主体部分结构阐述
	The rest of this paper is structured as follows. 
	Sec. \ref{Sec-2} introduces the related works.
	Sec. \ref{Sec-3} formally presents the problem formulation of CP over packet loss channels and gives the system model.
	Then, Sec. \ref{Sec-4} elaborates on
	the proposed SoM-MTM framework. 
	Sec. \ref{Sec-5} shows details of our experimental setup.
	Further, Sec. \ref{Sec-6} provides comprehensive results to demonstrate the efficacy of SoM-MTM. 
	Finally, Sec. \ref{Sec-7} concludes the paper and discusses future perspectives.
	
	\section{Related Work}\label{Sec-2}
	\subsection{Data Transmission for Cooperative Perception}\label{SecWork-1}
	Some works in the field of CP focus on efficient data transmission, moving beyond pure compression under ideal communication assumptions \cite{V2VNet,Where2comm}. 
	%For instance, 
	RoCooper \cite{RoCooper} and V2X-INCOP \cite{V2X-INCOP} respectively leverage mutual information from other viewpoints and historical moments, to cope with communication degradation or interruption, 
	%thereby enhancing the robustness of 
	under specific cooperation modes. 
	\cite{LCRN} proposes LCRN, which can repair the received feature under packet loss channels solely by exploiting the intrinsic correlations within perceptual features themselves, making it more generic.
	Recent studies have begun to integrate multi-agent interaction with channel conditions. \cite{Self-Adaptive} and \cite{Afformer} achieve selective feature fusion through adaptive strategies, attaining gains under different channel qualities. Coop-WD \cite{Coop-WD} regards recovery from corrupted information as a generative process and utilizes diffusion models to generalize across different levels of interference. 
	However, most of them mainly focus on object detection in V2X scenarios and are closely coupled with specific detection backbones, limiting their scalability to more general scenarios and applications.
	
	JSCC
	%-based semantic communication
	can empower the sensory data transmission by simplifying communication modules through error-tolerance abilities of deep neural networks.
	Many studies design JSCC-based transceiver systems to accomplish specific data transmission with low overhead, with visual images being a representative example.
	DeepJSCC \cite{DeepJSCC} is among the earliest works to realize wireless image compression and reconstruction. 
	Subsequent works \cite{ViT-JSCC-MIMO,SwinJSCC,SoM-DCAT,GAN-VAE-JSCC,DiT-JSCC} have incorporated advanced models into image codecs such as Vision Transformer \cite{ViT}, Swin Transformer \cite{Swin}, and even generative models like Generative Adversarial Networks (GAN) \cite{GAN}.
	
	A series of efforts have also shown that JSCC can bypass complete data reconstruction and support other specific downstream tasks, including classification \cite{recovery_and_classification,class_dc,multi-task-1}, re-identification and retrieval \cite{ReID,retrieval}, detection and segmentation \cite{Importance_Segmentation,3D_LiDAR_JinShi,SoM-MIMO}.
	Such works fundamentally exploit the end-to-end fitting
	capability of neural networks, and can indeed obtain joint gains under specific tasks and transmission configurations.
	For example, \cite{SoM-DCAT} considers image reconstruction under imperfect channel state information; \cite{class_dc} jointly designs classification and digital modulation; and \cite{SoM-MIMO} exploits MIMO precoding to enhance instance segmentation. 
	
	%However, generality and scalability remain key challenges.
	However, such an end-to-end design and optimization principle restricts their generalization and scalability, making them difficult to work well once the channel configuration, perception task, or even dataset changes.
	Although some studies have explored multi-task systems, they are essentially multi-branch extensions of existing architectures \cite{multi-task-1,multi-task-2}.
	Besides, their neural architectures are usually attached to mature perception backbones with inconsistent communication modeling assumptions, resulting in limited interpretability.
	
	Against this backdrop, we seek to develop a more general-purpose visual feature interaction framework that transcends mere fitting to specific tasks and configurations.
	Different from existing JSCC-based approaches,
	our proposed framework serves as a plug-and-play enhancement module over packet loss channels, without being bound to specific underlying configurations.
	%之前的建模优点2
	Instead, it uniformly abstracts them at the application layer, making the modeling more universal, requiring much fewer modifications to existing systems, which means lower deployment costs.
	%之前的建模优点3
	Moreover, the transceiver design is relatively decoupled and is not limited to point-to-point communication, which can flexibly support various cooperation modes such as multicasting and random multiple access.

	\subsection{Masked Image Modeling}\label{SecWork-2}
	Masked image modeling has emerged as a powerful self-supervised paradigm for visual representation learning. As the most representative, MAE \cite{MAE} can reconstruct images from a subset of visible patches, forcing the model to learn global semantic structures and contextual correlations from incomplete observations. 
	Through such perceptual contextual learning, MAE can be effectively transferred to downstream tasks such as classification and segmentation, while exhibiting strong scalability across different model sizes.
	
	\begin{figure*}[!t]
		\centering
		\includegraphics[width=0.9\linewidth]{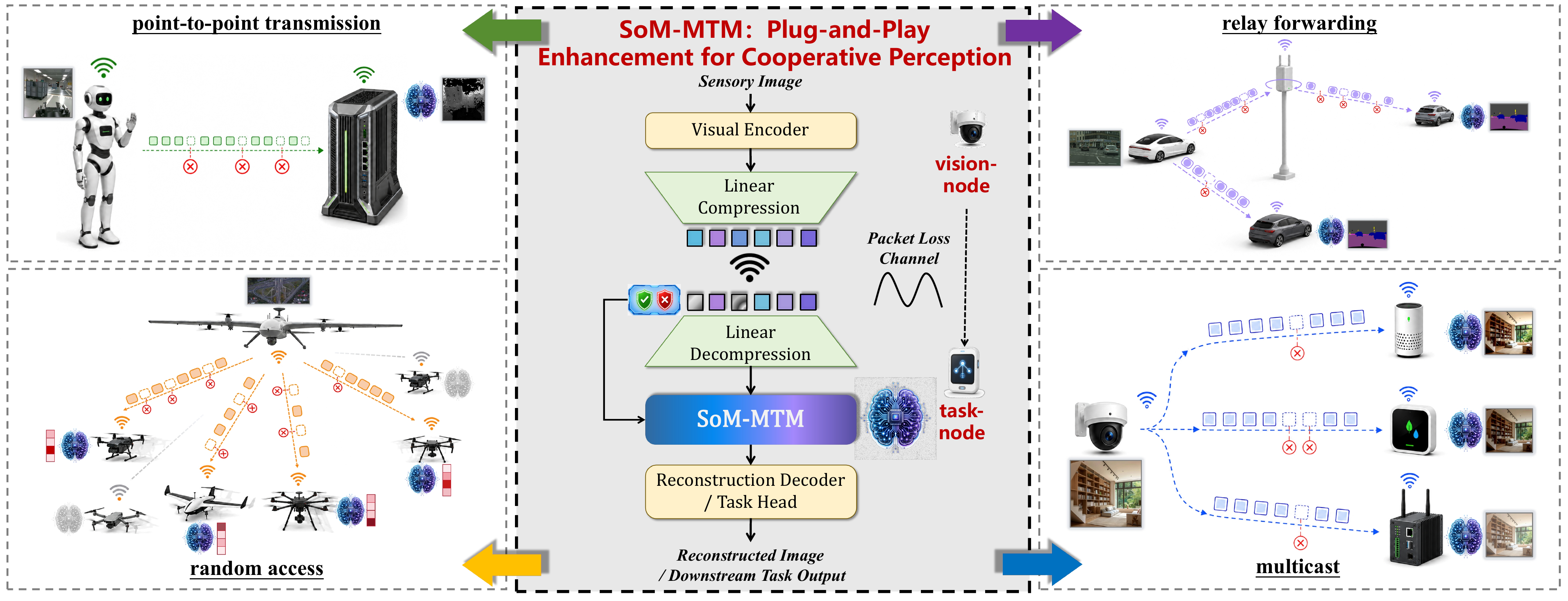}
		\caption
		{	
			Illustration of the proposed SoM-MTM plug-and-play framework for cooperative perception data transmission and its application scenarios.
		}
		\label{fig_overall}
	\end{figure*}
	
	%其他MIM方法
	Following MAE, other MIM methods improve visual contextual learning through different reconstruction targets and masking strategies. 
	BEiT \cite{BEiT} introduces higher-level token prediction to enhance semantic supervision, SimMIM \cite{SimMIM} simplifies the pipeline by directly predicting raw pixels with a lightweight head, and MaskFeat \cite{MaskFeat} further validates the effectiveness of feature-level prediction targets via hand-crafted visual features as supervision.
	They consistently demonstrate the capability of MIM to learn semantic dependencies and recover missing information from partial observations.
	
	%总结句
	These approaches emphasize visual representation and are used for image pre-training and downstream understanding, which are not directly applicable to data transmission. 
	Meanwhile, as discussed in Sec. \ref{Sec-1}, masking can be viewed not only as a manually designed training strategy, but also as packet validation status information over the lossy channels, indicating their potential connection. 
	Therefore, adapting MIM to communication-efficient transmission is both necessary and promising for empowering CP with plug-and-play visual representation capabilities.
	
	\section{System Model and Problem Formulation}\label{Sec-3}
	As depicted in Fig. \ref{fig_overall}, SoM-MTM is integrated into the CP framework in a plug-in manner to enhance feature transmission over packet loss channels.
	The agents participating in CP can be divided into the \textbf{vision-node} and \textbf{task-node}. The former collects the real-time sensory images and efficiently shares them with the task-node, while the task-node agent either reconstructs perceptual information through a decoder or directly performs a downstream perception task through a lightweight task head, depending on the specific requirement.
	
	In this process, the combination of the visual encoder and the reconstruction decoder or task head inherently constitutes an existing perception model, which can be simply split and deployed.
	\textbf{SoM-MTM operates at the task-node as a plug-and-play enhancement module}, refining the received information and enabling existing models to be readily adapted to multi-agent cooperation. 
	It is worth emphasizing that, under the packet loss-based information sharing paradigm, the numbers of vision-node and task-node agents are not restricted to a one-to-one relationship. 
	As shown in Fig. \ref{fig_overall}, the vision-node can multicast to multiple task-node agents; when the task-node is far away or the link condition is limited, data packets can also be forwarded through relay nodes.
	
	Building on the above, our system focuses on environmental perception tasks based on a single $k$-channel image input $\bm{S} \in \mathbb{R}^{H_0 \times W_0 \times k}$ , where $\mathbb{R}$ denotes the set of real numbers, $H_0$ and $W_0$ denote the height and width of the image.
	The cooperation process can be divided into three sequential steps:

	\subsubsection{Feature Extraction and Compression at the Vision-Node}
	The input image $\bm{S}$ is first fed into the visual encoder to extract compact perceptual features. 
	Modern visual image processing is increasingly built upon Vision Transformer (ViT) \cite{ViT} and its variants, where the image is tokenized with patches as basic units, while semantic correlations are modeled through self-attention to capture high-level semantics:
	\begin{equation}   
		\label{vision_encoder}
		\bm{F}=\mathrm{Enc}_\alpha(\bm{S})   
	\end{equation}
	where $\bm{F} \in \mathbb{R}^{N \times D}$ denotes the feature token sequence with length $N$ and channel dimension $D$. 
	To avoid ambiguity with the transmission channel or physical channel,
	we refer to the channel here as “feature-channel”.
	Subsequently, to enable efficient transmission and save communication resources, we compress its dimensionality through a simple fully connected (FC) layer to extract its principal components:
	\begin{equation}   
		\label{fc_encoder}
		\bm{Z}=\mathrm{FC}_{\sigma_1}(\bm{F})   
	\end{equation}
	where $\bm{Z} \in \mathbb{R}^{N \times d}$ is the compressed token sequence with the same length as $\bm{F}$ but a lower feature-channel dimension $d$, which is proportional to the size of a single data packet.
	The Compression Ratio (CR) that characterizes the efficiency of data transmission can be defined as:
	\begin{equation}   
		\label{CR_define}
		\text{CR}=\frac{N \times d}{H_0 \times W_0 \times k} 
	\end{equation}
	
	\subsubsection{Token-Level Packet Transmission over Packet Loss Channel}
	One of our core principles is \textbf{“token as packet”}, where each data packet exactly corresponds to a latent token with dimension $d$. Under this encapsulation pattern, each token is either perfectly transmitted or suffers from severe corruption and distortion, due to the layered redundancy coding and error correction mechanisms in existing protocols.
	Before token-by-token transmission, \textbf{interleaving coding} is performed to increase the spatial randomness of packet loss distribution:
	\begin{equation}   
		\label{interleaving}
		\bm{Z}^{\text{int}}=\bm{\Pi}({\bm{Z}})=[\bm{Z}_{\pi(1)},\bm{Z}_{\pi(2)},…,\bm{Z}_{\pi(N)}] 
	\end{equation}
	where random interleaving operation $\bm{\Pi}$ only permutes the token order, and the order indices are  predefined between the vision-node and the task-node.
	
	The data packet $\bm{\mathrm{TxData}}_i$ corresponding to the $i$-th token ($1 \le i \le N$), it is observed as $\bm{\mathrm{RxData}}_i$ at the task-node
	after undergoing a series of communication procedures. Through the redundancy-based decoding function $\mathcal{D}_{\mathrm{pkt}}$, the recovered token sequence and status information are then obtained as:
	\begin{equation}
		\label{receive_and_verify}
		(\bm{\widetilde{Z}},\bm{M})=\bm{\Pi}^{-1}\left(
		\mathcal{D}_{\mathrm{pkt}}
		\left(
		\{\bm{\mathrm{RxData}}_i\}_{i=1}^{N}
		\right)
		\right)
	\end{equation}
	where $\widetilde{\bm{Z}} \in \mathbb{R}^{N \times d}$ denotes the received token sequence after de-interleaving, and $\bm{M} \in\{0,1\}^{N}$ is the packet status mask:
	\begin{equation}
		\label{M_define}
		M_i =
		\begin{cases}
			1, & \text{indicating that }\widetilde{\bm{Z}}_i == \bm{Z}_i \text{ is ensured}\\
			0, & \text{otherwise}
		\end{cases}
	\end{equation}
	The packet loss rate is defined as the expected proportion of zero entries $R_L = 1 - \mathbb{E}\left[\frac{1}{N}\sum_{i=1}^{N}M_i\right]$,
	%\begin{equation}
	%	R_L = \mathbb{E}\left[\frac{1}{N}\sum_{i=1}^{N}(1-m_i)\right]
	%\end{equation}
	which is jointly determined by the wireless environment quality and the performance of communication modules, such as precoding, channel equalization, and symbol detection.
	
	Our modeling is uniformly abstracted at the application layer, regardless of underlying configurations.
	Therefore, without loss of generality, we assume that the channel state information (CSI) during transmission of the $i$-th token remains constant as $\bm{H}_i$, and the channel transmission is given by:
	\begin{equation}
		\label{basic_channel}
		\bm{Y}_i = \bm{H}_i \cdot \bm{X}_i+\bm{N}_i
	\end{equation}
	where $\bm{X}_i$ and $\bm{Y}_i$ denote the transmitted and received complex symbols corresponding to the token, and $\bm{N}_i$ is the electromagnetic noise. 
	For tractable simulation, we suppose that
	when the channel quality is sufficiently high, it can be interacted without error, as the equation below:
	\begin{equation}
		\label{NS_condition}
		M_i==1 
		\Longleftrightarrow 
		\left\|\bm{H}_i\right\| \geq \psi
	\end{equation}
	By adjusting the threshold $\psi$, we can simulate variations in the underlying modules and cover different loss rates.
	
	\subsubsection{SoM-MTM Enhancement at the Task-Node}
	Based on the key insight from \textbf{“packet loss as masking”}, SoM-MTM regards perfectly interacted and lossy data packets as visible and masked features, and leverages an MIM-style visual model for enhancement.
	First, different from the vision-node, we need to restore the feature dimension to $D$:
	\begin{equation}   
		\label{fc_decoder}
		\widetilde{\bm{F}}=\mathrm{FC}_{\sigma_2}(\widetilde{\bm{Z}})
	\end{equation}
	Similar to the MAE decoder, we utilize a learnable masked token $\bm{z}^{\text{mask}} \in \mathbb{R}^{D}$ to fill the lossy positions of the received feature as $\overline{\bm{F}}^{\text{in}}=\widetilde{\bm{F}} \cdot {\bm{M}} + \bm{z}^{\text{mask}} \cdot (1-\bm{M})$.
	\begin{comment}
		\begin{equation}   
			\label{fill}
			\overline{\bm{F}}=\widetilde{\bm{F}} \cdot {\bm{M}} + \bm{z}^{\text{mask}} \cdot (1-\bm{M})
		\end{equation}
	\end{comment}
	Building on this, our SoM-MTM can support pluggable feature refinement via perceptual contextual learning, formulated as:
	\begin{equation}   
		\label{SoM-MTM-eq}
		\overline{\bm{F}}^{\text{ref}}=\bm{\mathrm{MTM}}_\theta(\overline{\bm{F}}^{\text{in}},\bm{M})
	\end{equation}
	Different from MIM approaches, its input consists not only of feature $\overline{\bm{F}}$, but also of additional packet validation status $\bm{M}$, which can embody another innovation of our design: \textbf{“masking as prior”}.
	
	Finally, according to the requirement, the image $\hat{\bm{S}}$ is reconstructed or the result $\bm{O}$ for a specific task is output as:
	\begin{equation}
		\label{final_out}
		\begin{cases}
			\hat{\bm{S}}= \mathrm{Dec}_\beta(\overline{\bm{F}}^{\text{ref}}),\quad
			&\text{for general reconstruction} \\
			\bm{O}= \mathrm{Head}_\gamma(\overline{\bm{F}}^{\text{ref}}),\quad 
			&\text{for downstream task}
		\end{cases}
	\end{equation}
	
	Integrating all above, our goal is to employ SoM-MTM as a plug-and-play module and jointly optimize it with the remaining parts of the model, so as to achieve optimal performance under the given CR and packet loss rate $R_L$ constraints:
	\begin{equation}
		\label{total_problem_define}
		\begin{split} %\mathcal{P}
		\max_\Lambda \quad \bm{Perf}
		(\hat{\bm{S}} / \bm{O}),\quad
		\Lambda=&{\{\theta,\bm{z}^{\text{mask}},\sigma_1,\sigma_2,\alpha,\beta/\gamma\}} \\
		\mathrm{s.t.} \quad
		\text{CR} \leq \text{CR}^{\text{given}},&\;
		R_L = R_L^{\text{given}}
		\end{split}
	\end{equation}
	
	\section{SoM-Driven Masked Token Model Design}\label{Sec-4}
	In order to achieve efficient plug-and-play visual feature refinement and enhancement under packet loss channel conditions, we propose a SoM-driven Masked Token Model \textbf{(SoM-MTM)} framework.
	In this section, we detail its design principles guided by the SoM philosophy of communication-perception intelligent integration.
	We select Swin Transformer (Swin) as the backbone that utilizes window-based attention for perceptual contextual learning. 
	On this foundation, we take the packet status information as crucial prior knowledge, innovatively introduce external routing to improve the MoE mechanism.
	For training efficiency, we leverage transfer learning to optimally balance knowledge from multiple aspects.
	
	\subsection{Network Structure and Overall Workflow}\label{SecMethod-1}
	\begin{figure*}[!t]
		\centering
		\includegraphics[width=0.92\linewidth]{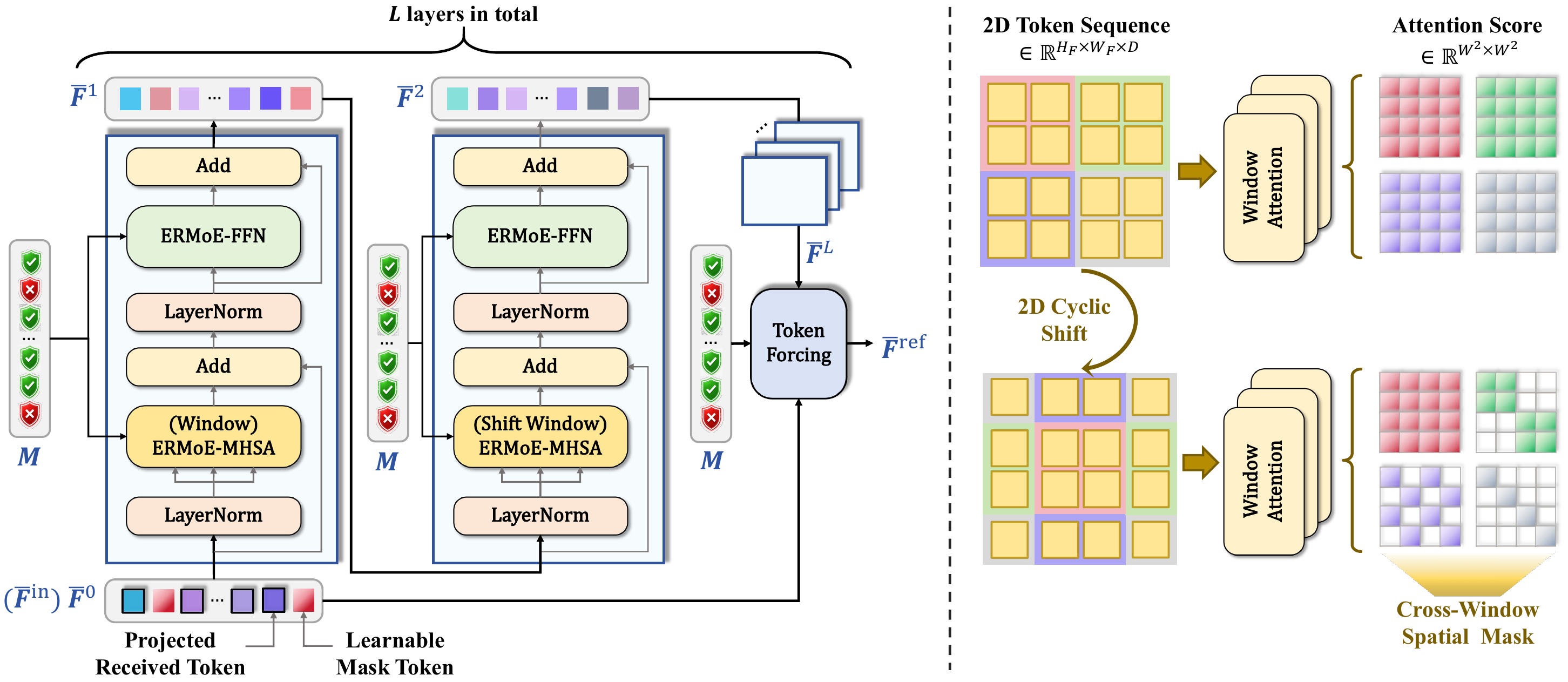}
		\caption
		{	
			Overall network architecture of SoM-MTM, our proposed perceptual feature context learning and refinement scheme based on Swin Transformer.
		}
		\label{fig_swin}
	\end{figure*}

	Building upon conventional natural language processing (NLP)-oriented attention \cite{All_you_need} mechanisms, ViT \cite{ViT} emphasizes aligning the processing paradigm of visual images as token-based sequence modeling.
	Swin \cite{Swin} further considers the requirements of fine-grained environmental perception, enabling more efficient and targeted local feature extraction through Window Multi-Head Self-Attention (W-MHSA) and Shifted Window Multi-Head Self-Attention (SW-MHSA). Therefore, we adopt \textit{2D Swin} as the backbone network, whose characteristics are highly compatible with our requirements of regional contextual learning.
	The overall architecture of the scheme is presented in Fig. \ref{fig_swin}.
	
	Specifically, the model is composed of $L$ stacked \textit{2D Swin} layers, where each layer computes self-attention within windows of a predefined size $W$.
	They uniformly divide the token sequence in a non-overlapping manner, which can reduce computational complexity and introduce an appropriate inductive bias: \textbf{perceptual contextual learning relies on spatially adjacent features}. The self-attention follows the format below:
	\begin{equation}
		\label{QKV_attention}
		\text{Attention}(\bm{Q},\bm{K},\bm{V})=\text{Softmax}(\frac{\bm{Q}\bm{K}^T}{\sqrt{D_H}}+\bm{P}_r+\bm{M}_s)\bm{V}
	\end{equation}
	where $\bm{Q},\bm{K},\bm{V} \in \mathbb{R}^{W^2 \times D_H}$ are the Query, Key and Value matrices.
	Multi-head attention employs $H$ parallel heads, each operating on a subspace of dimension $D_H$ where $D_H=D/H$.
	$\bm{P}_r$ and $\bm{M}_s$ are two bias terms added to the attention scores, representing the learnable relative positional encoding and the spatial isolation mask for cyclic shifts, respectively.
	
	As shown in Fig. \ref{fig_swin}, to introduce inter-window connection, the effective association regions are redefined between consecutive layers by shifting the 2D windows. Specifically, the boundary regions with the width or height of $\lfloor \frac{W}{2} \rfloor$ are cyclically shifted, while $\bm{M}_s$
	is used to prevent improper attention introduced while maintaining efficient parallel computation. 
	Therefore, W-MHSA and SW-MHSA are performed alternately across layers, and two successive \textit{2D Swin} blocks can be illustrated as follows:
	\begin{equation}
		\label{Swin Transformer}
		\begin{split} 
			&\hat{\bm{F}}^{l}=\text{LN}(\text{W-MHSA}(\overline{\bm{F}}^{l-1},\bm{M}))+\overline{\bm{F}}^{l-1}\\
			&\overline{\bm{F}}^{l}= \text{LN}(\text{FFN}(\hat{\bm{F}}^{l},\bm{M}))+\hat{\bm{F}}^{l} \\
			&\hat{\bm{F}}^{l+1}=\text{LN}(\text{SW-MHSA}(\overline{\bm{F}}^{l},\bm{M}))+\overline{\bm{F}}^{l}\\
			&\overline{\bm{F}}^{l+1}= \text{LN}(\text{FFN}(\hat{\bm{F}}^{l+1},\bm{M}))+\hat{\bm{F}}^{l+1} \\
		\end{split}
	\end{equation}
	where $\overline{\bm{F}}^{j}$ ($0 \le j \le L$) represent the output features of the $j$-th layer ($\overline{\bm{F}}^{0}=\overline{\bm{F}}^{\text{in}}$).
	The FFN (Feed Forward Neural Network) is used in conjunction with self-attention to enhance the feature-correlation representation within each token. 
	
	After $L$ layers of progressive regional contextual learning, the refined representation of the complete perceptual features $\overline{\bm{F}}^{L}$ is obtained.
	An additional operation similar to the pre-training process of MAE is required, where the successfully received tokens before refinement are directly copied over, referred to as \textbf{“token forcing”}: 
	\begin{equation}
		\label{token_forcing}
		\overline{\bm{F}}^{\text{ref}}=\overline{\bm{F}}^{\text{in}} \cdot \bm{M} + \overline{\bm{F}}^{L} \cdot (1-\bm{M})
	\end{equation}
	Here, $\bm{M}$ indicates whether the token is successfully received or corrupted by packet loss.
	Token forcing strategy enhances the model interpretability, SoM-MTM explicitly leverages perfectly transmitted tokens to recover the contextual information. 
	Accordingly, the vision-node and task-node can collaborate more effectively to accomplish the cooperative perception task.
	\subsection{ERMoE Mechanism Design}\label{SecMethod-2}
	
	The \textit{2D Swin}-based architecture indeed provides a powerful foundation for perceptual contextual learning, its design principle is still inherited from conventional computer vision tasks, where all input tokens are treated equally and attention is established solely according to visual correlations. 
	However, the objective of SoM-MTM is not normal visual understanding, but contextual enhancement and refinement of perceptual features over packet loss channel conditions. 
	Under the circumstances, in order to enable adaptive
	capability for transmission conditions, we need to
	effectively utilize other relevant information beyond the image feature itself.
	
	To be specific, the packet status information $\bm{M}$ explicitly indicates whether each packet is successfully decoded according to Eq. (\ref{receive_and_verify}) or corrupted during transmission.
	Since it determines the reliability of each token and can be readily obtained, we embed it into the processing procedure as important prior information.
	We design an \textbf{External Routing Mixture-of-Experts methodology (ERMoE)}, which regards the packet mask as communication-side assistance to appropriately integrate information from different domains. As shown in Eq. (\ref{Swin Transformer}), we respectively introduce sparse and dense ERMoE into FFN and MHSA of \textit{2D Swin}, which are described separately below.
	
	%ERMoE-FFN
	We first elaborate on the module design of \textbf{ERMoE-FFN}.
	Mixture-of-Experts (MoE) \cite{MoE_ICLR} has emerged as an effective paradigm for handling token heterogeneity and expanding model capacity while maintaining computational efficiency.
	By using a lightweight routing network as a “dispatcher”, replacing the shared FFN with MoE has become a common technique in emerging large language models (LLM) and visual foundation models (FM). The idea of categorizing and processing heterogeneous data distributions in parallel is consistent with our objective to some degree.
	
\begin{figure}[!t] % 强制图片位置
		\centering
		% 上方子图 (a)
		\subfloat[]{
			\includegraphics[width=0.87\linewidth]{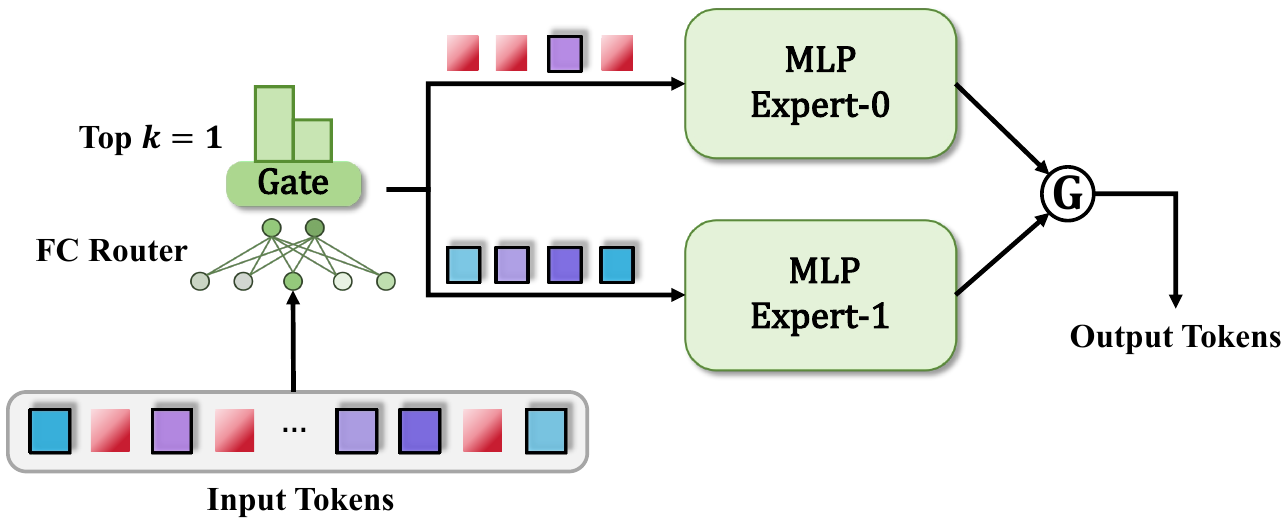} 
		}
		\vspace{0.01cm} % 调整上下子图间距
		\subfloat[]{
			\includegraphics[width=0.87\linewidth]{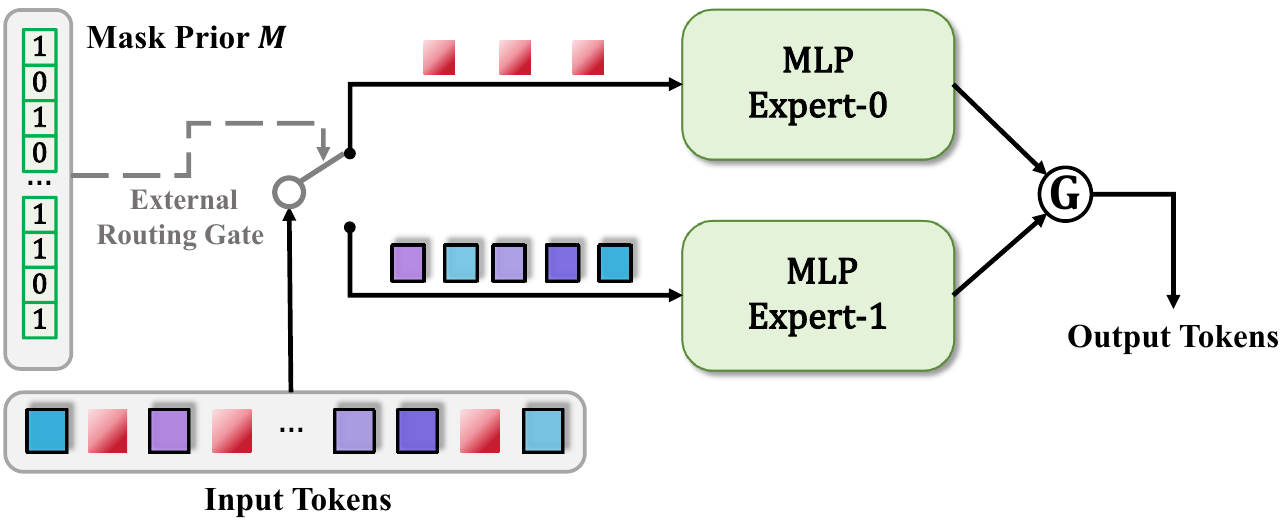} 
		}
		\caption{
			Comparison of expert selection in the FFN layer. (a) Standard sparse MoE based on token-dependent routing; (b) The proposed ERMoE mechanism based on prior mask information guiding.
		}
		\label{fig_ermoe_ffn}
	\end{figure}

	Meanwhile, the distribution discrepancy that SoM-MTM needs to handle mainly arises not from scenarios or tasks, but from different channel effects experienced by individual tokens, which distinguishes it from standard MoE. 
	We develop a two-expert ERMoE-FFN network, where only one is sparsely activated for each token. 
	Its structure and differences from standard MoE are depicted in Fig. \ref{fig_ermoe_ffn}.
	Specifically, successfully decoded tokens and corrupted tokens require different feature-processing strategies in the FFN layer, while such differences may be difficult to distinguish solely from their intrinsic distributions. Therefore, instead of learning routing decisions from token features, it is more reasonable to directly exploit the packet mask for expert selection. Taking the token $\hat{\bm{F}}_i$ with index $i$ as the input to the FFN layer, it can be formulated as:
	\begin{equation}
		\label{Eq_ERMoE_FFN}
		\text{ERMoE-FFN}(\hat{\bm{F}},\bm{M})[i] =
		\begin{cases}
			E_0(\hat{\bm{F}}_i), & \text{if} \quad M_i==0\\
			E_1(\hat{\bm{F}}_i), & \text{if} \quad M_i==1
		\end{cases}
	\end{equation}
	where $E_0$ and $E_1$ are both two-layer position-wise Multi-Layer Perceptron (MLP) experts. Since only one expert is sparsely activated, it includes no additional computational cost. 
	Moreover, such external routing mechanism can even further improve efficiency by facilitating model parallelization.
	
	\begin{figure*}[!t] % 强制图片位置
		\centering
		\includegraphics[width=0.92\linewidth]{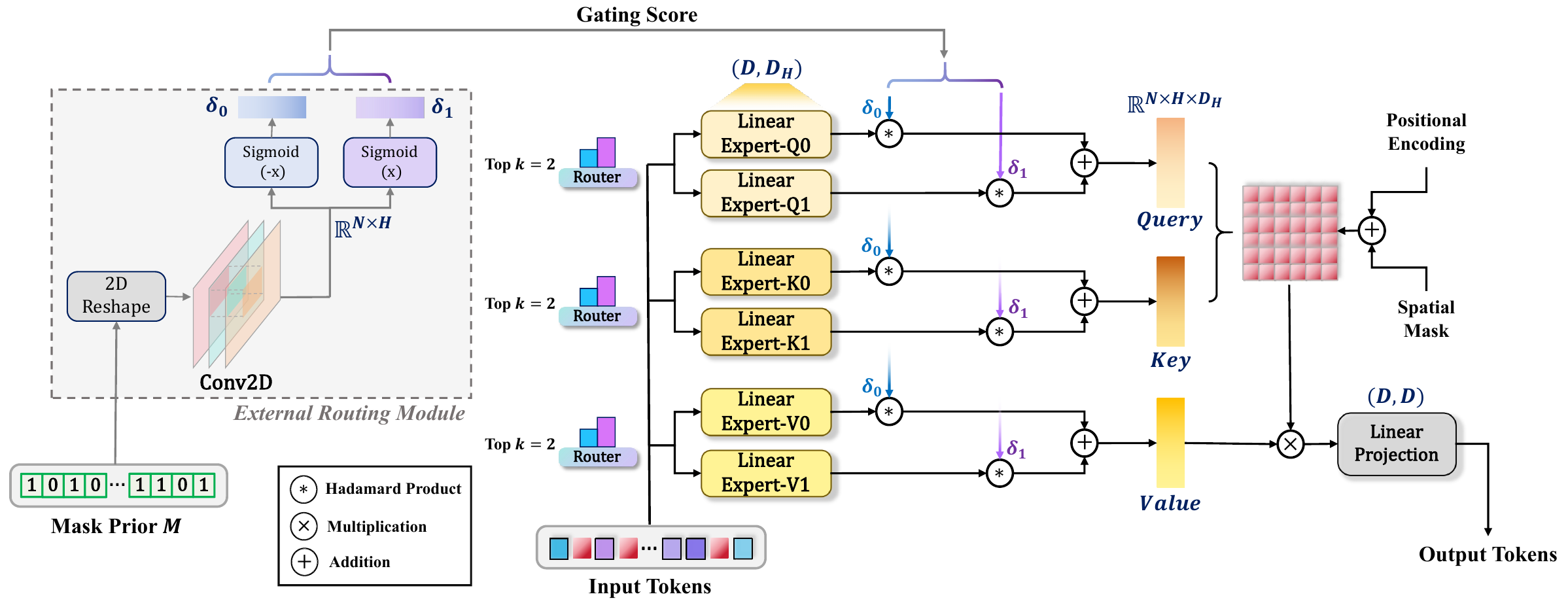} % 实际图片路径
		\caption{
			Diagram of the ERMoE-MHSA modules.
			The Query, Key, and Value embeddings are each produced by two densely activated experts, whose contributions are modulated by shared gating scores derived from the packet mask information $\bm{M}$.
		}
		\label{fig_ermoe_mhsa}
	\end{figure*}

	%ERMoE-MHSA
	As for \textbf{ERMoE-MHSA}, the situation is slightly different, whose module diagram can be seen from Fig. \ref{fig_ermoe_mhsa}.
	The primary role of MHSA is to establish contextual associations among tokens and aggregate information from neighboring regions. 
	During contextual information aggregation, packets with different reliability should contribute differently to attention formation. 
	In other words, the attention weights from successfully decoded and distorted tokens should inherently be different, which also requires explicit information as auxiliary guidance.
	
	Therefore, densely activated ERMoE is introduced into the MHSA layer.
	This design is conceptually related to Mixture-of-Heads (MoH) \cite{MoH},
	as they both enhance attention modeling through expertized projections.
	ERMoE-MHSA employs external routing to extend the isolated linear layer into two experts ($\mathrm{Top}\text{-}k=2$ with different weights).
	Taking the input token $\overline{\bm{F}}$ associated within the $h$-th attention head as an example, the $\bm{Q}/\bm{K}/\bm{V}$ matrices in Eq. (\ref{QKV_attention}) are expressed by:
	\begin{equation}
		\label{Eq_ERMoE_MHSA}
		\begin{cases}
			\bm{Q}=\bm{\delta_0}[:,h] \cdot \overline{\bm{F}}\bm{W}_Q^0+\bm{\delta_1}[:,h] \cdot \overline{\bm{F}}\bm{W}_Q^1\\
			\bm{K}=\bm{\delta_0}[:,h] \cdot \overline{\bm{F}}\bm{W}_K^0+\bm{\delta_1}[:,h] \cdot \overline{\bm{F}}\bm{W}_K^1\\
			\bm{V}=\bm{\delta_0}[:,h] \cdot \overline{\bm{F}}\bm{W}_V^0+\bm{\delta_1}[:,h] \cdot \overline{\bm{F}}\bm{W}_V^1\\
		\end{cases}
	\end{equation}
	with learnable matrices $\bm{W}_Q^0,\bm{W}_Q^1,\bm{W}_K^0,\bm{W}_K^1,\bm{W}_V^0,\bm{W}_V^1 \in \mathbb{R}^{D \times D_H}$.
	They serve as linear experts to control the computation of contextual associations among tokens within the window, and further the generation of attention scores. 
	While $\bm{\delta_0},\bm{\delta_1} \in \mathbb{R}^{N \times H}$ and denote the gating scores generated by the \textbf{External Routing Module}:
	\begin{equation}
		\label{External_Routing}
		\begin{split} 
			 \bm{x} &=\mathrm{Conv2D}(\bm{M}_{2D})
			\\
			 \bm{\delta_0},\bm{\delta_1} &=\sigmoid(-\bm{x}_{1D}),\sigmoid(\bm{x}_{1D})
		\end{split}
	\end{equation}
	Instead of directly using the binary packet mask, a lightweight size-preserving layer $\mathrm{Conv2D}$ is adopted to exploit the spatial distribution patterns of packet loss and generate smooth gating scores.
	With the assistance of such external routing factors, contextual associations can be adaptively adjusted according to the packet reception status, enabling more flexible and targeted perceptual feature learning and fusion.
	
	To conclude, we enhance both feature-level refinement and context-level association with the help of ERMoE. 
	By replacing implicit feature learning with explicit routing gates, prior information is leveraged to substantially reduce process uncertainty, leading to better model interpretability. 
	It can also fundamentally avoid issues like expert imbalance and routing collapse \cite{Switch}, thereby making contextual learning more stable. 
	In addition, compared with standard MoE, the external routing mechanism is easier to parallelize so that it is more friendly to both training and deployment.
	
	\subsection{Model Training}\label{SecMethod-3}
	As a plug-and-play model, SoM-MTM operates at the task-node to achieve perceptual feature enhancement and refinement under packet loss channels. To ensure that SoM-MTM can coordinate well with the other components of the entire model and achieve optimal cooperative perception ability, we adopt a joint training strategy.

	Inspired by transfer learning, the overall training process is divided into two progressive stages, as described below:
	\begin{list}{--}{
			\setlength{\leftmargin}{1.5em}
			\setlength{\itemsep}{0.1em}
		}
		\item \textit{Stage \ding{172} (Ideal Communication Training)}:
		In the first stage, we focus on the cooperative perception task under perfect communication conditions, where the packet loss is temporarily assumed to be absent.
		Specifically, during this stage, we bypass the proposed SoM-MTM along with the FC compression and decompression ($\sigma_1$ and $\sigma_2$) module.
		This stage aims to exclude the influence of multi-agent cooperation and enable the model to acquire basic environmental perception capability. Once the parameters of the visual encoder and the decoder/task head have largely converged, they are transferred to the next stage as a favorable initialization.
		\item \textit{Stage \ding{173} (SoM-MTM Joint Training with Packet Loss)}:
		Building upon the model above, we proceed to the second stage by activating all relevant parameters including our SoM-MTM, (i.e., $\Lambda$ in Eq. (\ref{total_problem_define}))  to form the complete pipeline.
		At this stage, SoM-MTM is plugged in for feature contextual learning and refinement over packet loss channels. Meanwhile, the other components of the model are also updated, allowing the vision-node to provide more effective contextual association information and the task-node to perform more robustly under distorted conditions.
	\end{list}
	Such a “step-by-step” learning strategy allows the network to focus on knowledge acquisition at each stage, thereby alleviating the performance degradation caused by directly learning difficult tasks in an end-to-end manner.
	
	At the same time, to further enhance model interpretability, we define the training loss function as follows:
	\begin{equation}   
		\label{total_loss}
		Loss=L_{p}+\omega \cdot \text{MSE}(\overline{\bm{F}}^{\text{ref}},\overline{\bm{F}}^{\text{gt}})
	\end{equation}
	where $L_{p}$ represents the constraint of the perception task itself, while the second term constrains the feature recovery effect of SoM-MTM. $\overline{\bm{F}}^{\text{gt}}$ denotes the ground truth under perfect interaction, which can be obtained by: $\overline{\bm{F}}^{\text{gt}}=\mathrm{FC}_{\sigma_2}(\bm{Z})$.
	$\omega$ is the balancing weight and we set it to $0.1$.
	
	Through this training process, the proposed SoM-MTM can better operate within the overall framework in a plug-and-play manner, fully exploiting its capability for contextual representation learning and realizing efficient enhancement for cooperative perception systems.
	 
	\section{Experimental Setup}\label{Sec-5}
	\subsection{Tasks and Performance Metrics for Validation}\label{SecSetup-1}
	To validate the effectiveness of SoM-MTM, we conduct extensive evaluations across different tasks and configurations based on RGB images ($k\sneq3$), with the specific experimental logic outlined as follows.
	
	First, we test the fundamental capabilities of SoM-MTM based on general environmental information reconstruction. In this scenario, the task-node requires the decoder to decompress the feature and increase the spatial resolution. 
	Sec. \ref{SecExp-1} to \ref{SecExp-3} present comparisons from various perspectives to highlight the comprehensive superiority of SoM-MTM.
	We select the peak signal-to-noise ratio (PSNR) as the evaluation metric at this stage.
	Higher PSNR values indicate better image reconstruction quality, as defined by:
	\begin{equation}
		\label{PSNR-define}
		\text{PSNR}(\bm{S},\bm{\hat{S}}) = 10 \log_{10} \frac{(\max \bm{S})^2}{\text{MSE}(\bm{S},\bm{\hat{S}})} \text{ (dB)}
	\end{equation}
	where $\max \bm{S}$ denotes the maximum possible value of the original image $\bm{S}$ (255 for the 8-bit color).
	%During training, ${\text{MSE}(\bm{S},\bm{\hat{S}})}$  serves appropriately as the loss function.
	
	Beyond above, we also explore whether SoM-MTM can be directly applied to downstream perception tasks. 
	Sec. \ref{SecExp-4} shows the numerical evaluations for two specific tasks, image classification and semantic segmentation. For image classification, Top-1 Accuracy is adopted as the metric:
	\begin{equation}
		\label{ACC-define}
		\text{Top-1 Acc} = \frac{1}{N_s}\sum_{i=1}^{N_s}(f(\bm{S}_i)==label_i)
	\end{equation}
	where $N_s$ denotes total sample size. 
	As for semantic segmentation, the metric is the mean intersection over union (mIOU) of $N_{cls}$ categories:
	\begin{equation}
		\label{mIOU-define}
		\text{mIOU} = \frac{1}{N_{cls}}\sum_{i=1}^{N_{cls}}\frac{P \cap G}{P \cup G}
	\end{equation}
	where $P$ is the set of pixel regions predicted for a certain category, and $G$ is the actual set for this category.
	Both metrics indicate better performance when higher values are achieved.  
	%The cross-entropy between the predicted label vector and the ground truth one-hot vector is adopted as the loss function.
	
	\subsection{Datasets and Channel Condition Settings}\label{SecSetup-2}
	For general image reconstruction, we train the model on PLACES365 \cite{PLACES365}, a highly diverse collection rich in various scenes. Due to the inherently self-supervised nature of this task, we also directly evaluate its zero-shot performance on other unseen datasets. 
	For labeled classification and segmentation tasks, we train and test the model on the NWPU-RESISC45 \cite{NWPU45} and NYUv2 \cite{NYUv2} datasets, respectively.
	During training, all images are resized into the shape of $256\times256$.
	
	Meanwhile, we adopt the widely used channel generator QuaDRiGa \cite{QuaDRiGa} to simulate the real-time channel of concern corresponding to Eqs. (\ref{basic_channel})(\ref{NS_condition}).
	We generate multiple sets of channel CSI evolutions, selecting one for each sample. At each discrete time step, one packet is transmitted, and the packet loss rate ($R_L$) can be adjusted by setting thresholds. The packet arrival rate ($R_A=100 \%-R_L$) during training ranges between 30\% and 80\%.
	\begin{table*}[!t]
		\centering
		\renewcommand{\arraystretch}{1.2} % 行高增加50%
		\caption
		{
			Model configurations of different perception tasks for experimental validation.
		}
		\label{tab:tasks}
		% 子表
		\begin{tabular}{@{}!{\vrule width 1.2pt}c !{\vrule width 1.2pt} c|c|c!{\vrule width 1.2pt}c|c|c!{\vrule width 1.2pt}@{}}
			\specialrule{1.2pt}{0pt}{0pt}
			{\textbf{Task}} & Metric & %Training 
			Perception Loss $L_p$ & Training Dataset & Visual Encoder & Reconstruction Decoder&Task Head\\
			\specialrule{1.0pt}{0pt}{0pt}
			\textbf{Reconstruction} &PSNR&${\text{MSE}(\bm{S},\bm{\hat{S}})}$& PLACES365& Swin& Swin&\\
			\hline
			\textbf{Classification} &Top-1 Acc&Cross Entropy& NWPU-RESISC45& Swin& &Lightweight ViT\\
			\hline
			\textbf{Segmentation} &mIOU&Cross Entropy& NYUv2& Heavyweight ViT %(initialized from SAM Encoder) 
			& &CNN+FC\\
			\specialrule{1.2pt}{0pt}{0pt}
		\end{tabular}
	\end{table*}
	
	\subsection{Benchmarks}\label{SecSetup-3}
	To evaluate the effectiveness of our proposed SoM-MTM, we compare it against other distinct baseline schemes.
	We first include the following methods, which are also plug-and-play.
	They are applicable to all evaluated tasks:
	\begin{itemize}
		\item \textbf{LCRN \& LCRN*} \cite{LCRN}: LCRN  utilizes a multi-scale convolutional and residual connection network architecture to repair intermediate features damaged by lossy channels. Unlike our modeling, its data organization format treats a feature dimension as the basic unit of data packets.
		%, rather than a token representing a feature pixel.
		For a more comprehensive and fair comparison, we experiment with both data organization formats. Hereafter, LCRN refers to the version consistent with our modeling, where a token serves as the basic data packet unit, while LCRN* represents its originally proposed format, where each dimension of features is the basic unit.
		\item \textbf{Vanilla MAE} \cite{MAE}: Conventional MIM methods like MAE possess capabilities in contextual learning and data completion. We treat features after packet loss as “pseudo-images” masked by MAE and fill the masked tokens utilizing an MAE decoder based on the vanilla ViT. Its processing architecture is consistent with ours, making the comparison more effective in highlighting the advantages of our network module design.
	\end{itemize}

	%To ensure fairness, all the ready-to-use schemes including SoM-MTM rely on the same encoder and decoder/head architectures, sharing identical training process of the first stage.
	
	%As for generic image reconstruction, considering that the decoder architecture itself is based on token-level attention and possesses the ability to learn and recover distorted features, we further introduce the following benchmarks for comparison:
	Further, for generic image reconstruction, we additionally consider baselines which are specifically applicable to it. 
	This is attributed to that
	reconstruction-oriented encoder-decoder architectures naturally exploit token-level attention and possess the ability to learn and recover distorted features. 
	Incorporating these benchmarks provides a more rigorous evaluation of the model scalability advantage of our SoM-MTM.
	
	\begin{itemize}
		\item \textbf{Swin-S} \cite{SwinJSCC}: SwinJSCC is the SOTA solution for image compression and reconstruction. Due to differences in channel modeling, we remove its adaptation modules. Here, we employ its small model (Swin-S) as \textbf{an ablation version for all others.} It serves as a lower bound to verify whether other designs can genuinely achieve gains.
		\item \textbf{Swin-B \& Swin-B*} \cite{SwinJSCC}: SwinJSCC-Base (Swin-B) Version employs more transformer layers in the encoder and decoder to achieve stronger fitting capabilities. Comparative analysis with it can highlight the cost efficiency advantages. Similarly, 
		%in the subsequent descriptions, 
		Swin-B and Swin-B* denote the versions where tokens and feature-channels, respectively, serve as the basic unit of data packets.
	\end{itemize}

	\subsection{Network Configuration}\label{SecSetup-4}
	\begin{table}[!t]  %超参表格
		\centering
		\caption{The hyper-parameters for network training.}
		\label{tab:hyper-parameters}
		\renewcommand{\arraystretch}{1.2}
		\setlength{\tabcolsep}{6pt}
		\begin{tabular}{@{}cc@{}}
			\specialrule{1.0pt}{0pt}{0pt} % 加粗线
			\textbf{Config} & \textbf{Value} \\
			\specialrule{1.0pt}{0pt}{0pt} % 加粗线
			Batch size & $16$ \\ \hline
			\multirow{2}{*}{Epochs} & $60$ for \textit{Stage \ding{172}}\\ 
			& $80$ for \textit{Stage \ding{173}}\\ \hline
			Optimizer & AdamW ($\beta_1=0.9$,$\beta_2=0.999$) \\ \hline
			Learning rate schedule & MultiStepLR (gamma=$[0.5,0.5]$) \\ \hline
			\multirow{2}{*}{Milestones} & $[30,50]$ for \textit{Stage \ding{172}}\\ 
			& $[40,65]$ for \textit{Stage \ding{173}}\\ \hline
			Base learning rate & $1\times10^{-4}$ \\
			\specialrule{1.0pt}{0pt}{0pt} % 加粗线
		\end{tabular}
	\end{table}
	For different perception tasks, the detailed training configurations and architectural choices of the model are summarized in Table \ref{tab:tasks}. 
	For the reconstruction task, the dimension $D$ of the extracted feature, is set to $320$. For downstream classification and segmentation, this dimension is uniformly set to $256$. The CR relative to the original image is set to $1/12$ for all tasks.
	The numbers of self-attention heads and Transformer blocks are set to $H=8$ and $L=6$, respectively.
	
	Table \ref{tab:hyper-parameters} details the hyperparameter settings for the network training phase. We adopt the two-stage training strategy mentioned in Sec. \ref{SecMethod-3}.
	Specifically, for semantic segmentation, the visual encoder is initialized with pre-trained parameters from visual foundation model Segment Anything Model (SAM) \cite{SAM}, and we fine-tune it with a learning rate scaled by $0.01$.
	All models are implemented with PyTorch $2.7$, with four NVIDIA GeForce RTX4090 GPUs.
	
	\section{Performance Evaluation and Discussions}\label{Sec-6}
	
	\subsection{Numerical Results Comparison with Baselines}\label{SecExp-1}
	\subsubsection{Basic In-domain Comparative Study} 
	\begin{figure}[!t] % 强制图片位置
		\centering
		\includegraphics[width=0.9\linewidth]{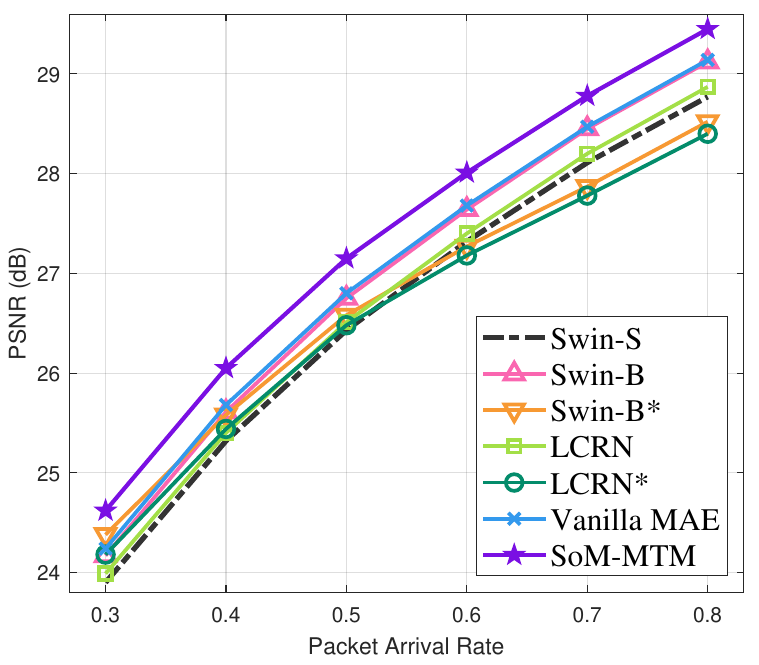} % 实际图片路径
		\caption{Comparison of PSNR performance between the proposed SoM-MTM and baselines under trained channel conditions ($30 \% \le R_A \le 80 \%$).}
		\label{pic_main_inner}
	\end{figure}
	
	We first test the in-distribution fitting capability for the generic self-supervised task. After pre-training on the large-scale PLACES365 dataset, Fig. \ref{pic_main_inner} illustrates the reconstruction performance of different methods within the trained arrival rate range ($30 \%$ to $80 \%$).
	It is evident that our proposed SoM-MTM consistently maintains optimal performance, achieving at least an 8\% improvement (0.34 dB) over other baselines.
	Particularly taking Swin-S as the reference, which serves as the shared theoretical lower bound, such improvement is stable and significant.
	
	While the effectiveness of others is inconsistent, particularly Swin-B* and LCRN*, where data organization based on feature-dimensions as the unit even underperforms the theoretical lower bound at certain arrival rates. This validates the rationality of our modeling, which regards tokens as the fundamental unit, offering greater flexibility and adaptive ability to dynamic channel conditions.
	
	\begin{table}[!b]
		\centering
		\caption{
			Evaluation of complexity and computation cost per batch of each model. The \textbf{boldface} denotes the highest value.}
		\label{tab:cost}
		\renewcommand{\arraystretch}{1.35}
		\setlength{\tabcolsep}{5pt}
		\resizebox{\linewidth}{!}{%
			\begin{tabular}{@{}c|ccccc@{}}
				\specialrule{0.9pt}{0pt}{0pt} % 加粗线
				\textbf{Model} & \textbf{Swin-S} & \textbf{Swin-B(*)} & \textbf{LCRN(*)} & \textbf{Vanilla MAE} & \textbf{SoM-MTM} \\
				\specialrule{0.9pt}{0pt}{0pt} % 加粗线
				Parameters & 11.96 M& 18.28 M& \textbf{42.48 M} & 19.36 M& 26.13 M\\ \hline
				FLOPs & 420.1 G& \textbf{523.6 G}& 427.2 G& 450.3 G& 457.9 G\\ \hline
				Inference Time &12.7 ms&\textbf{18.9 ms}&15.3 ms&14.9 ms&17.8 ms\\
				\specialrule{0.9pt}{0pt}{0pt} % 加粗线
			\end{tabular}
		}
	\end{table}
	\begin{table*}[!t]
		\centering
		\caption
		{
			The zero-shot performance on different unseen image datasets, averaged the metrics across all tested channel conditions.
		}
		\label{tab:zero_shot}
		\renewcommand{\arraystretch}{1.2}
		\setlength{\tabcolsep}{5pt}
		\begin{tabular}{@{}ccc!{\vrule width 1.1pt}ccccccc@{}}
			\specialrule{1.1pt}{0pt}{0pt} % 加粗线
			\multicolumn{3}{c!{\vrule width 1.1pt}}{Dataset} & \multicolumn{7}{c}{PSNR Performance} \\ \hline
			Name & Description & Image Size & \underline{\textbf{Swin-S}} & \textbf{Swin-B} & \textbf{Swin-B*} & \textbf{LCRN} & \textbf{LCRN*} & \textbf{Vanilla MAE}& \textbf{SoM-MTM} \\
			\specialrule{1.1pt}{0pt}{0pt} % 加粗线
			\textbf{Mini-ImageNet} \cite{Mini_Imagenet} & Close-up of Objects & $128^2$ & \underline{26.19} & +0.07 & +0.11 & / & / & +0.57 & \textbf{+1.01} \\ \hline
			\textbf{NYUv2} \cite{NYUv2}& Indoor Home Environments & $256^2$ & \underline{26.32} & +0.10 & -0.01 & +0.14 & -0.34 & +0.71 & \textbf{+1.02} \\ \hline
			\textbf{RoboMIND} \cite{Robomind}& Robotic Workshops & $256^2$ & \underline{32.85} & +0.10 & +0.38 & +0.05 & +0.07 & +0.72 & \textbf{+1.27} \\ \hline
			\textbf{FFHQ} \cite{FFHQ}& Human Faces & $256^2$ & \underline{28.41} & +0.01 & +0.23 & +0.07 & +0.12 & +0.51 & \textbf{+0.87} \\ \hline
			\textbf{DAIR-V2X} \cite{Dair-v2x}& Vehicular Networks (V2X) & $768^2$ & \underline{31.51} & -0.02 & +0.00 & +0.01 & -0.25 & -4.69 & \textbf{+1.14} \\ \hline
			\textbf{DIV2K} \cite{DIV2K}& High-definition Artistic Images & $1024^2$ & \underline{27.29} & -0.01 & -0.06 & +0.02 & -0.25 & -4.15 & \textbf{+0.70} \\
			\specialrule{1.1pt}{0pt}{0pt} % 加粗线
		\end{tabular}
	\end{table*}
	\subsubsection{Model Cost and Efficiency}
	Building on the above, we further evaluate the “cost-effectiveness” of our SoM-MTM, i.e., the additional computational and storage costs corresponding to the aforementioned gains.
	The model size, floating point operations (FLOPs), and inference time can indicate the cost-effectiveness.  
	Thanks to its high-efficiency architecture, SoM-MTM achieves the optimal performance without significant extra storage and computational overhead, as shown in Table \ref{tab:cost}.
	Swin-S, serving as the ablated version, yields the lowest metrics across all evaluations. 
	Notably, LCRN \& LCRN* exhibit the highest parameter count because of high-dimensional convolutions with multi-layer skip connections and 
	Swin-B \& Swin-B* incur the maximum computational FLOPs and inference time owing to deeper network structure. 
	SoM-MTM does not achieve the highest value across all metrics, thus its cost remains entirely within acceptable limits.
	Although more layers of Transformers are included, the efficiency remains highly impressive thanks to the rational module arrangement and the model parallelism adopted by our designed ERMoE.
	
	\subsubsection{Out-of-Distribution Generalization}
	
\begin{figure}[!t] % 强制图片位置
		\centering
		\hspace*{\fill}
		\subfloat[Worse Channel Conditions]{
			\includegraphics[width=0.47\linewidth]{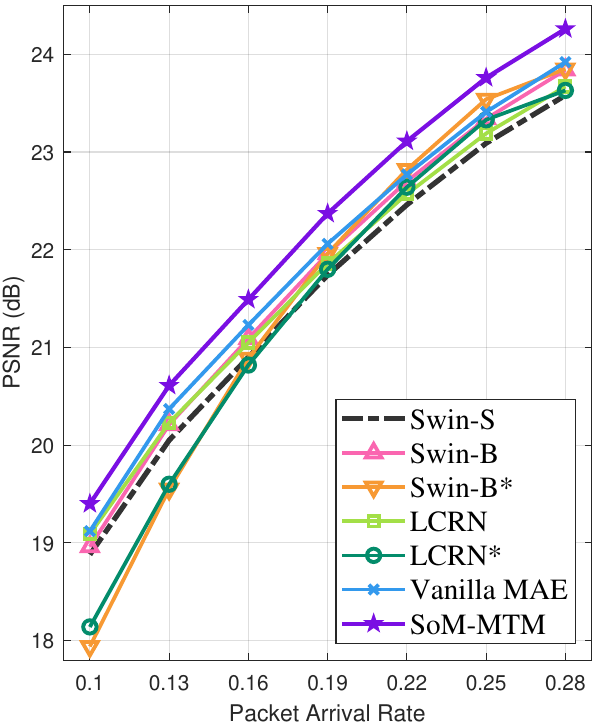} % 实际图片路径
			\label{pic-out-a}
		}
		\subfloat[Better Channel Conditions]{
			\includegraphics[width=0.47\linewidth]{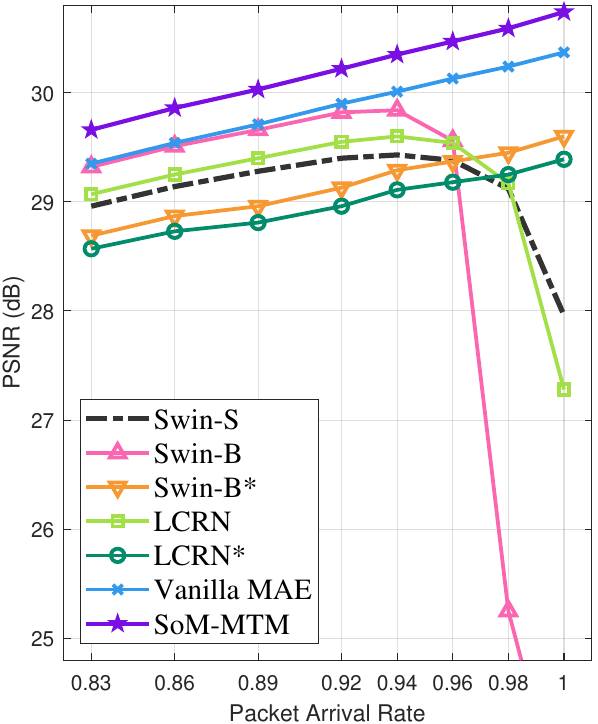} % 实际图片路径
			\label{pic-out-b}
		}
		\hspace*{\fill}
		\caption{Comparison of PSNR performance between SoM-MTM and baselines under unseen channel conditions (a) $R_A<30 \%$. (b) $R_A>80 \%$.}
		\label{pic_main_outer}
	\end{figure}

	This part involves directly performing zero-shot inference with the pre-trained models under distribution shift to validate the generalization capacity. First, Fig. \ref{pic_main_outer} illustrates the extrapolation ability of different schemes under varying channel conditions on the PLACES365 dataset, where \ref{pic-out-a} and \ref{pic-out-b} represent lower and higher packet arrival rates respectively.
	
	From the curves, it is evident that the generalizability of baselines is insufficient.
	When channel conditions deteriorate ($R_A$ decreases from $30 \%$), the performance of Swin-B* and LCRN*, which rely on feature-by-feature packet transmission, declines sharply, indicating weak robustness.
	More strikingly, when channel conditions are more favorable than that during training ($R_A>80 \%$), the result of Swin-S, Swin-B, and LCRN exhibits abnormal degradation. Moreover, the better the in-distribution performance, the more pronounced the decline. This reveals that they are sensitive to specific distributions during training and fail to learn universal knowledge of CP in this scenario, which is an extremely serious issue.
	In contrast, SoM-MTM consistently maintains the best ability under channel condition extrapolation.
	%Vanilla MAE that shares a similar processing pipeline with ours, retains suboptimal performance, with the gap primarily attributed to specific network choices and detailed design differences.

	Besides, we explore the generalizability on various types of unseen perceptual image types. Table \ref{tab:zero_shot} provides detailed information on six untrained datasets, along with the zero-shot performance of different methods. 
	For clarity, Swin-S, which acts as the theoretical lower bound, shows its metric underlined. All other methods present the PSNR gain relative to it, tested across an arrival rate range of $10 \%$ to $100 \%$.
	\begin{table*}[!t]
		\centering
		\caption
		{
			Results of ablation experiments for working flow and network design on the PLACES365 Dataset.
		}
		\label{tab:ab_1234}
		\renewcommand{\arraystretch}{1.15}
		\setlength{\tabcolsep}{6pt}
		\begin{tabular}{@{}cccc!{\vrule width 1.1pt}ccc@{}}
			\specialrule{1.1pt}{0pt}{0pt} % 加粗线
			\multicolumn{4}{c!{\vrule width 1.1pt}}{SoM-MTM Design} & \multicolumn{3}{c}{PSNR Performance} \\ \hline
			Interleaving Coding & Token Forcing & ERMoE-MHSA & ERMoE-FFN & $R_A=30 \%$ & $R_A=50 \%$ & $R_A=80 \%$  \\
			\specialrule{1.0pt}{0pt}{0pt} % 加粗线
			\ding {51} & \ding {51} & \ding {51} & \ding {51} & \textbf{24.62} & \textbf{27.15}  & \textbf{29.45}   \\ \hline
			\ding {51} & \ding {51} & \ding {51} & & 24.55 & 27.07 & 29.41 \\ \hline
			\ding {51} & \ding {51} &  &  & 24.50 & 27.01 & 29.37  \\ \hline
			\ding {51} &  & & & 24.40 & 26.90 & 29.21  \\ \hline
			& &  & & 23.33 & 26.03 & 28.51  \\ 
			\specialrule{1.1pt}{0pt}{0pt} % 加粗线			
		\end{tabular}
	\end{table*}

	It is evident that other approaches suffer from significant generalization deficiencies, with performance gains severely diminished or even turning negative. Among them, LCRN and LCRN* are constrained by structural rigidity and fail to function on low-resolution images, while the previously suboptimal MAE performs poorly on high-resolution images. % due to its inflexible processing paradigm.
	Our method consistently achieves the highest value across all datasets, with an average improvement of 23\% (0.89 dB) over the second, which is quite remarkable.
	
	\subsection{Ablation Study}\label{SecExp-2}
	
	In this subsection, we conduct comprehensive ablation studies in order to verify the effectiveness of various specialized designs in the proposed SoM-MTM.
	
	Table \ref{tab:ab_1234} reveals the specific effects of key designs in our SoM-MTM, including the network modules and processing strategies towards packet loss channels. For the architectural components, we sequentially remove ERMoE-FFN and ERMoE-MHSA, replacing them with the original MLP and self-attention in the vanilla Transformer. Building on this, we further abandon the token forcing at the task-node and the interleaving strategy at the vision-node.
	
	As displayed in Table \ref{tab:ab_1234}, the full model consistently yields the highest PSNR value.
	Comparing it with the ablation models, we observe a gradual competence degradation, which aligns with our expectations.
	This confirms the necessity of our SoM-based design, as the proper embedding of communication prior information during cooperation can significantly contribute to the enhancement of perceptual performance.
	Fig. \ref{pic_vis} visualizes the reconstructed images of the first and last row, corresponding to the complete SoM-MTM design and the fully ablated version (termed \textbf{SoM-MTM-AB}), based on out-of-distribution images from Table \ref{tab:zero_shot}.
	We ensure a fair comparison by maintaining the same number of lossy tokens.%, though their spatial positions differ due to interleaving encoding. 
	It is visually evident that SoM-MTM achieves lower distortion. It does well in capturing geometric and texture information, recovering lossy tokens based on unmasked regions. 
	In contrast, SoM-MTM-AB without our specialized designs, exhibits weaker contextual comprehension ability, resulting in significant local blurring and shape distortion in the images.
	This is also supported by the MSE error metrics. Compared to the ablated version, the full SoM-MTM consistently reduces distortion in regions corresponding to lossy tokens across all samples, while also lowering errors in unmasked regions in most cases.
	Our design is compatible with native image processing neural networks and can enhance cooperation performance with minimal overhead.
	
	\begin{table}[!b]
		\centering
		\caption
		{
			Cost and performance comparison against standard MoE architecture at $R_A=30\%$.
		}
		\label{tab:ab_moe}
		\renewcommand{\arraystretch}{1.15}
		\setlength{\tabcolsep}{5pt}
		\begin{tabular}{!{\vrule width 1.0pt}c!{\vrule width 0.8pt}ccc!{\vrule width 0.8pt}c!{\vrule width 1.0pt}}
			\specialrule{1.0pt}{0pt}{0pt} % 加粗线
			\multirow{2}{*}{\textbf{Scheme}} & \multicolumn{3}{c!{\vrule width 1.1pt}}{Cost Metric} & \multirow{2}{*}{PSNR (dB)}  \\ \cline{2-4}
			& Parameters & FLOPs& Training Time &  \\
			\specialrule{1.0pt}{0pt}{0pt} % 加粗线
			\textbf{SoM-MTM} & 26.13 M & 457.9 G & 192 ms & \textbf{24.62} \\
			\specialrule{1.0pt}{0pt}{0pt} % 加粗线
			\textbf{MoE 1/2} & 26.14 M & 450.3 G & 225 ms & 24.41 \\ \hline
			\textbf{MoE 1/4} & 37.23 M & 450.4 G & 235 ms & 24.30 \\ \hline
			\textbf{MoE 1/8} & 66.82 M & 450.5 G & 320 ms & 24.25 \\ \hline
			\textbf{MoE 2/2} & 26.14 M & 478.0 G & 252 ms & 24.56 \\ \hline
			\textbf{MoE 2/4} & 37.23 M & 478.1 G & 257 ms & 24.55 \\ \hline
			\textbf{MoE 3/8} & 66.82 M & 505.8 G & 332 ms & 24.57 \\ 
			\specialrule{1.0pt}{0pt}{0pt} % 加粗线		
		\end{tabular}
	\end{table}
	Because the processing of ERMoE incurs higher costs compared to ablation baselines, the performance gain is difficult to assess to some degree. 
	To further verify the effectiveness, we conduct a more rigorous ablation experiment by comparing it with a standard MoE architecture. 
	Specifically, we replace the linear layers generating $\bm{Q}/\bm{K}/\bm{V}$ embeddings in MHSA and the MLP layers in the FFN with standard MoE, where routing decisions are made by the tokens themselves rather than external information as SoM-MTM. 
	The results are detailed in Table \ref{tab:ab_moe}, where each row labeled “MoE $N_{act}$/$N_{tot}$” indicates the activation of $N_{act}$ experts out of a total of $N_{tot}$.
	\begin{figure*}[!t]
		\centering
		% 设置表格列间距（控制组内图片的紧密程度，数值越小越近）
		\setlength{\tabcolsep}{2.5pt} 
		% 设置行距（控制图片与底部文字的距离）
		\renewcommand{\arraystretch}{0.6} 
		
		% 计算图片宽度：总宽度减去组间额外的间隔，再除以9张图
		% 这里假设组间额外间隔为 1.5em，你可以根据实际情况微调
		\newlength{\imgwidth}
		\setlength{\imgwidth}{\dimexpr (\linewidth - 3em - 12\tabcolsep)/9 \relax}
		
		\begin{tabular}{@{}ccccccccc@{}}
			% --- 第一行：组名 (A, B, C) ---
			% \multicolumn{3}{c}{...} 表示合并3列并居中显示组名
			\multicolumn{3}{c}{\textbf{$R_A=30\%$}} & \multicolumn{3}{c}{\textbf{$R_A=50\%$}} & \multicolumn{3}{c}{\textbf{$R_A=80\%$}} \\
			% --- 使用 \cmidrule 画出带间隔的下划线 ---
			% (r) 表示缩短右边，(l) 表示缩短左边，这样三段线之间会有非常精致的缝隙
			\cmidrule[1.2pt](r){1-3}\cmidrule[1.2pt](lr){4-6}\cmidrule[1.2pt](l){7-9} 
			
			\footnotesize \textbf{Original Image} & \footnotesize \textbf{SoM-MTM} & \footnotesize \textbf{SoM-MTM-AB} & 
			\footnotesize \textbf{Original Image} & \footnotesize \textbf{SoM-MTM} & \footnotesize \textbf{SoM-MTM-AB} &
			\footnotesize \textbf{Original Image} & \footnotesize \textbf{SoM-MTM} & \footnotesize \textbf{SoM-MTM-AB}  \\
			
			\cmidrule(r){1-3}\cmidrule(lr){4-6}\cmidrule(l){7-9}
			% --- 第1行：9张图片 ---
			% 组A的3张图
			\includegraphics[width=\imgwidth]{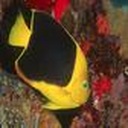} & 
			\includegraphics[width=\imgwidth]{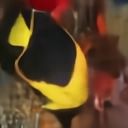} & 
			\includegraphics[width=\imgwidth]{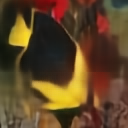} & 
			% 组B的3张图
			\includegraphics[width=\imgwidth]{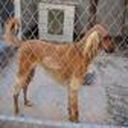} & 
			\includegraphics[width=\imgwidth]{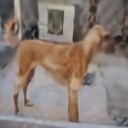} & 
			\includegraphics[width=\imgwidth]{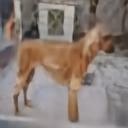} & 
			% 组C的3张图
			\includegraphics[width=\imgwidth]{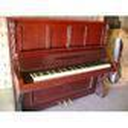} & 
			\includegraphics[width=\imgwidth]{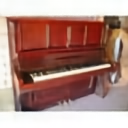} & 
			\includegraphics[width=\imgwidth]{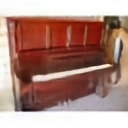} \\
			% --- 第1行：每张图底部的文字说明 ---
			\footnotesize Mini-ImageNet & \footnotesize 1.18 / \textbf{2.80} & \footnotesize \textbf{1.15} / 4.02 & 
			\footnotesize Mini-ImageNet & \footnotesize \textbf{1.01} / \textbf{2.45} & \footnotesize 1.14 / 2.69 &  
			\footnotesize Mini-ImageNet & \footnotesize \textbf{0.68} / \textbf{1.25} & \footnotesize 0.74 / 5.27 \\
			
			\cmidrule(r){1-3}\cmidrule(lr){4-6}\cmidrule(l){7-9}
			% --- 第2行：9张图片 ---
			% 组A的3张图
			\includegraphics[width=\imgwidth]{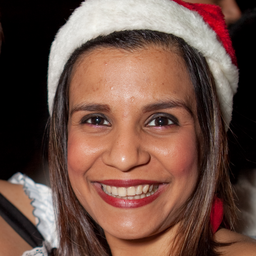} & 
			\includegraphics[width=\imgwidth]{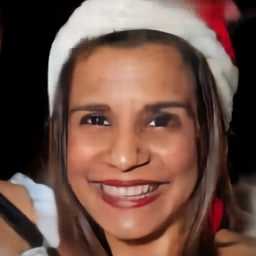} & 
			\includegraphics[width=\imgwidth]{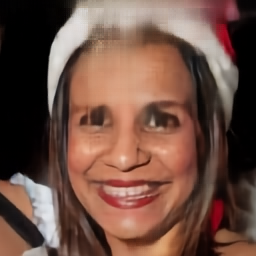} & 
			% 组B的3张图
			\includegraphics[width=\imgwidth]{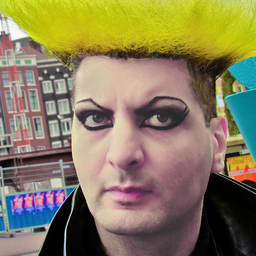} & 
			\includegraphics[width=\imgwidth]{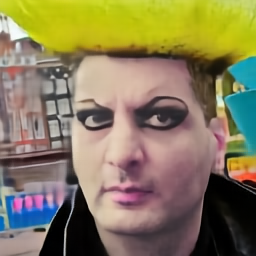} & 
			\includegraphics[width=\imgwidth]{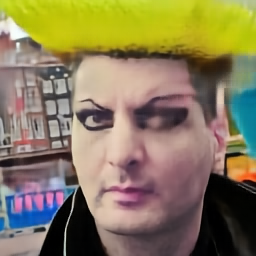} & 
			% 组C的3张图
			\includegraphics[width=\imgwidth]{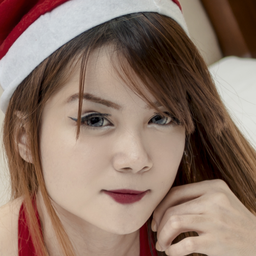} & 
			\includegraphics[width=\imgwidth]{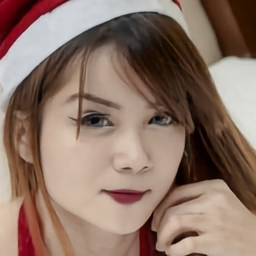} & 
			\includegraphics[width=\imgwidth]{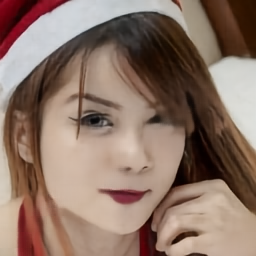} \\
			% --- 第2行：每张图底部的文字说明 ---
			\footnotesize FFHQ & \footnotesize 0.82 / \textbf{2.69} & \footnotesize \textbf{0.72} / 7.19 & 
			\footnotesize FFHQ & \footnotesize \textbf{1.47} / \textbf{4.09} & \footnotesize 1.92 / 5.88 &  
			\footnotesize FFHQ & \footnotesize \textbf{0.34} / \textbf{0.58} & \footnotesize 0.41 / 1.21 \\
			
			\cmidrule(r){1-3}\cmidrule(lr){4-6}\cmidrule(l){7-9}
			% --- 第3行：9张图片 ---
			% 组A的3张图
			\includegraphics[width=\imgwidth]{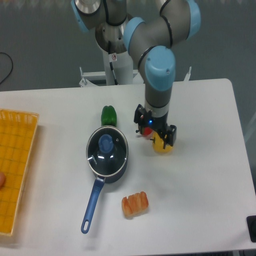} & 
			\includegraphics[width=\imgwidth]{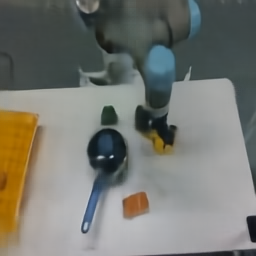} & 
			\includegraphics[width=\imgwidth]{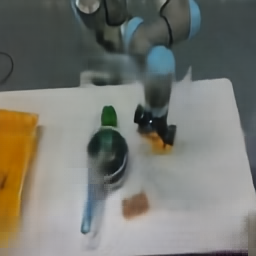} & 
			% 组B的3张图
			\includegraphics[width=\imgwidth]{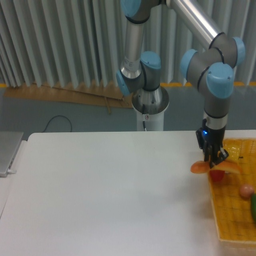} & 
			\includegraphics[width=\imgwidth]{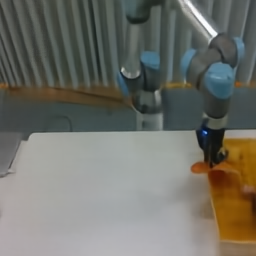} & 
			\includegraphics[width=\imgwidth]{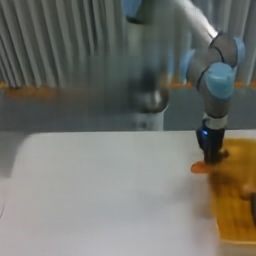} & 
			% 组C的3张图
			\includegraphics[width=\imgwidth]{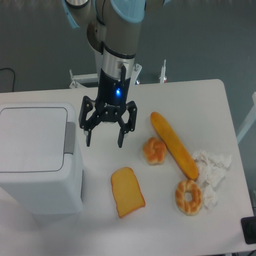} & 
			\includegraphics[width=\imgwidth]{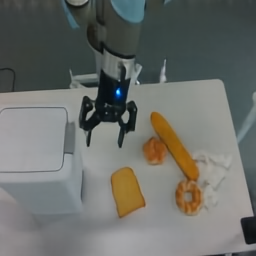} & 
			\includegraphics[width=\imgwidth]{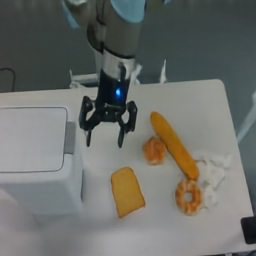} \\
			% --- 第3行：每张图底部的文字说明 ---
			\footnotesize RoboMIND & \footnotesize \textbf{0.24} / \textbf{1.50} & \footnotesize 0.26 / 2.51 & 
			\footnotesize RoboMIND & \footnotesize 0.19 / \textbf{0.46} & \footnotesize \textbf{0.18} / 1.66 &  
			\footnotesize RoboMIND & \footnotesize \textbf{0.14} / \textbf{0.19} & \footnotesize 0.18 / 0.32 \\
			
			\cmidrule(r){1-3}\cmidrule(lr){4-6}\cmidrule(l){7-9}
			% --- 第4行：9张图片 ---
			% 组A的3张图
			\includegraphics[width=\imgwidth]{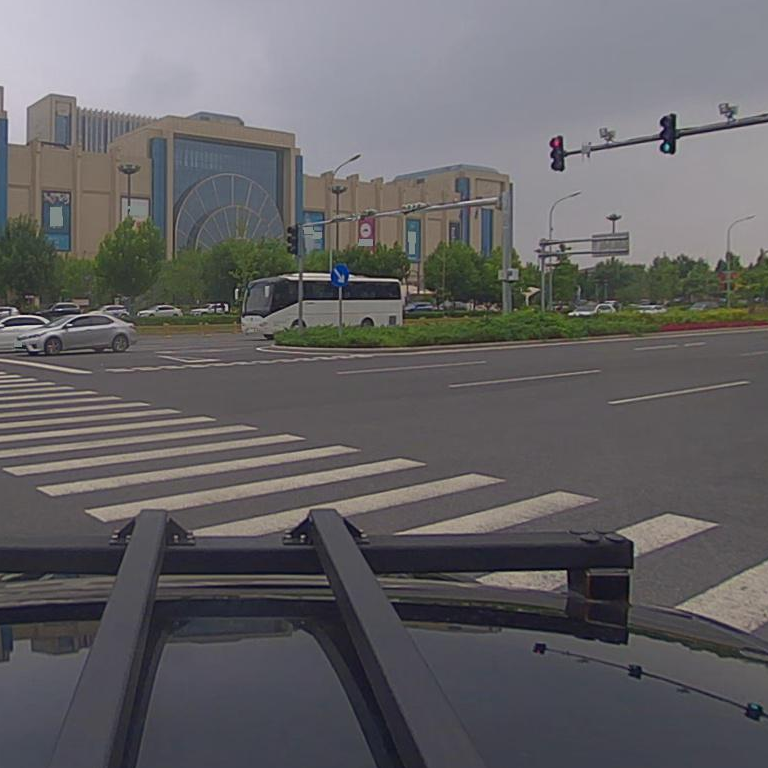} & 
			\includegraphics[width=\imgwidth]{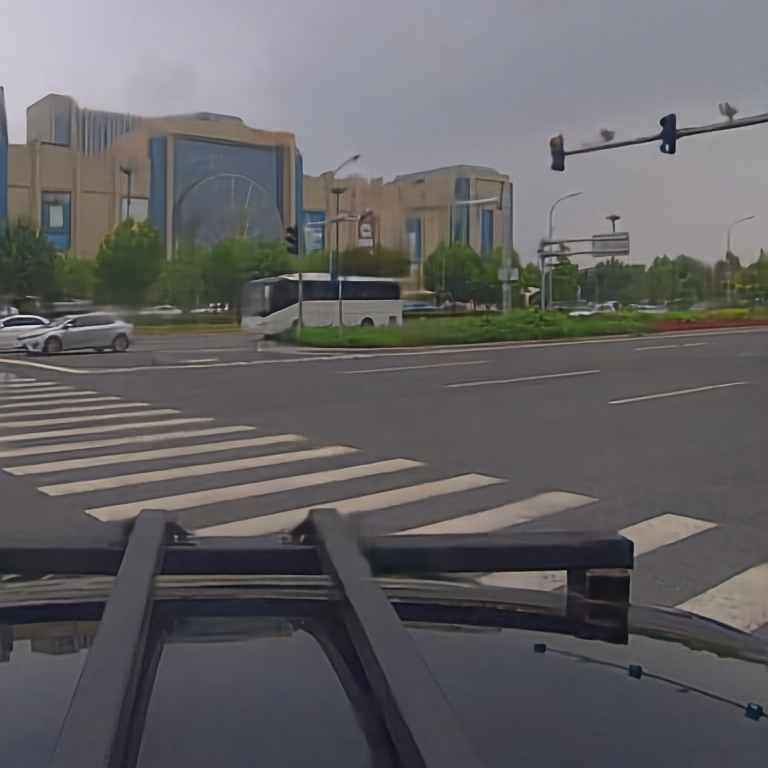} & 
			\includegraphics[width=\imgwidth]{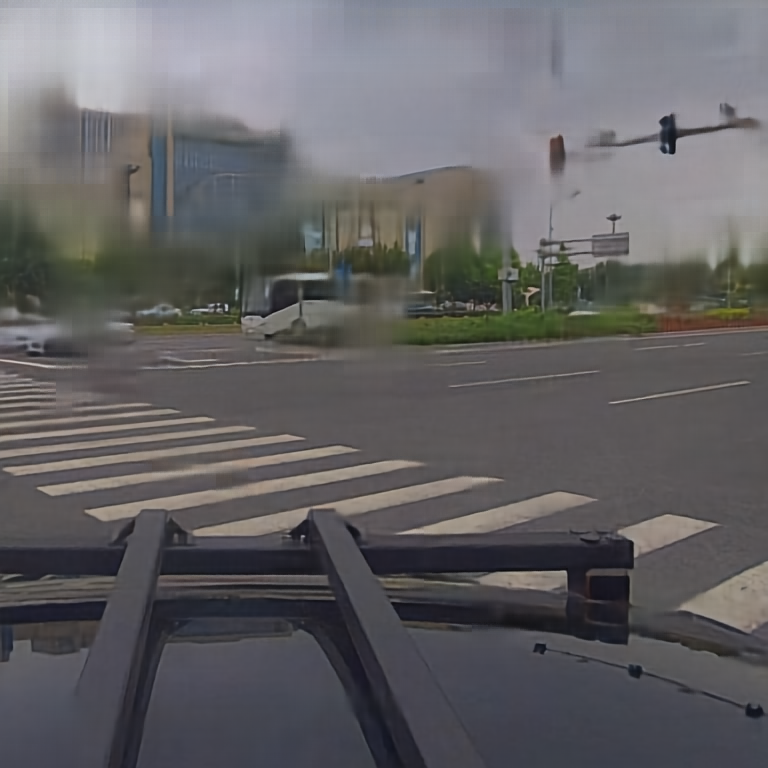} & 
			% 组B的3张图
			\includegraphics[width=\imgwidth]{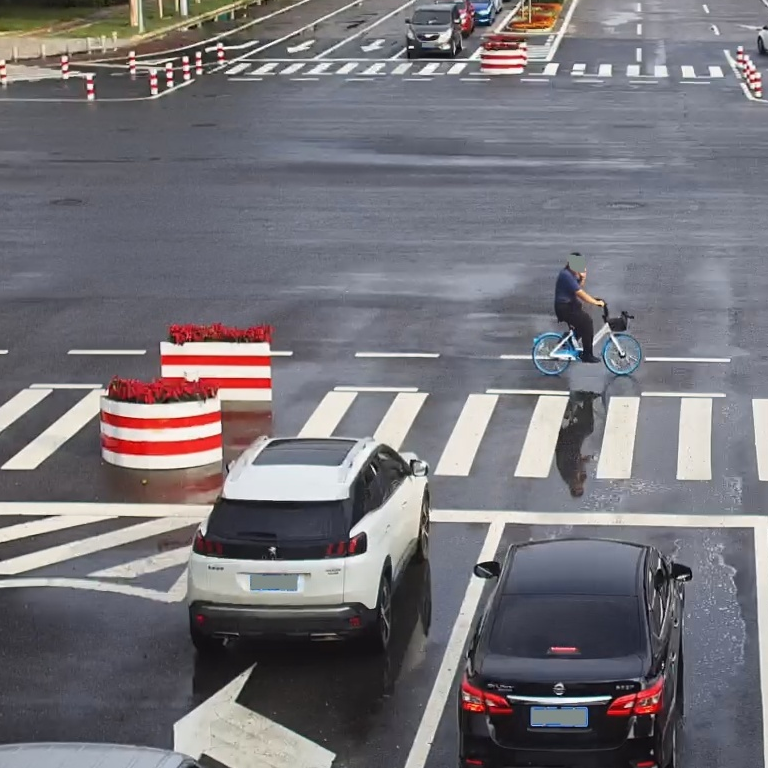} & 
			\includegraphics[width=\imgwidth]{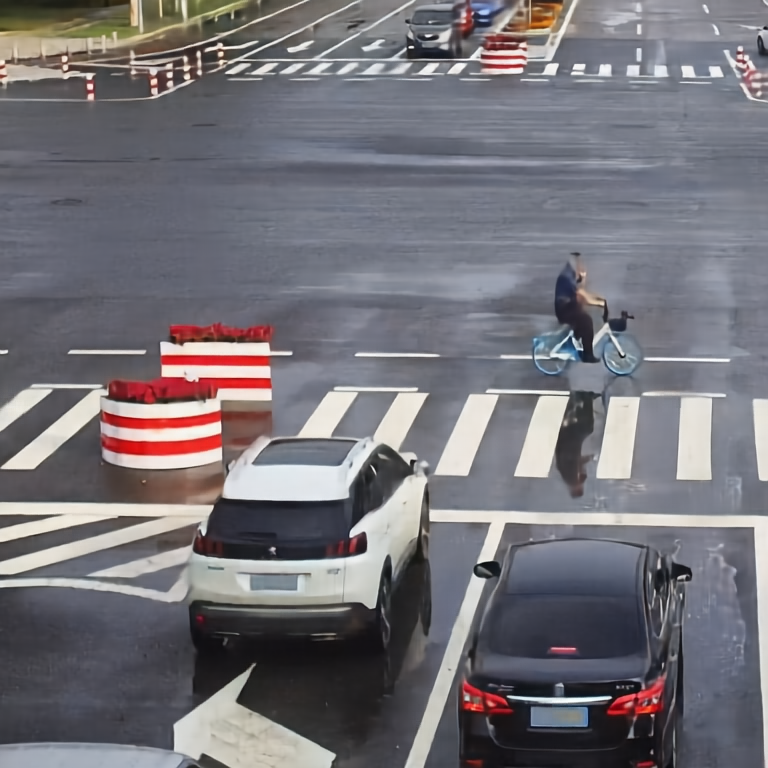} & 
			\includegraphics[width=\imgwidth]{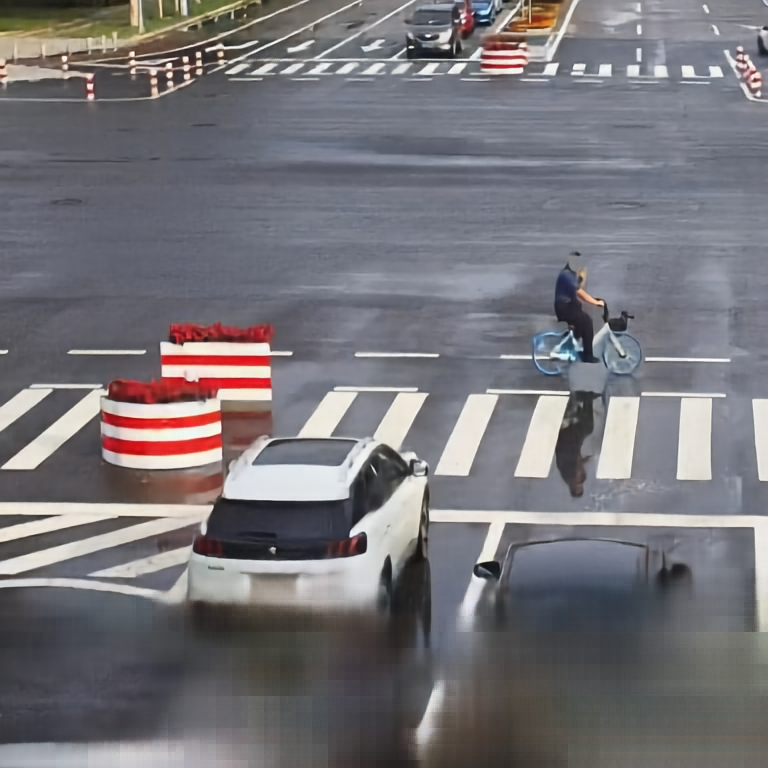} & 
			% 组C的3张图
			\includegraphics[width=\imgwidth]{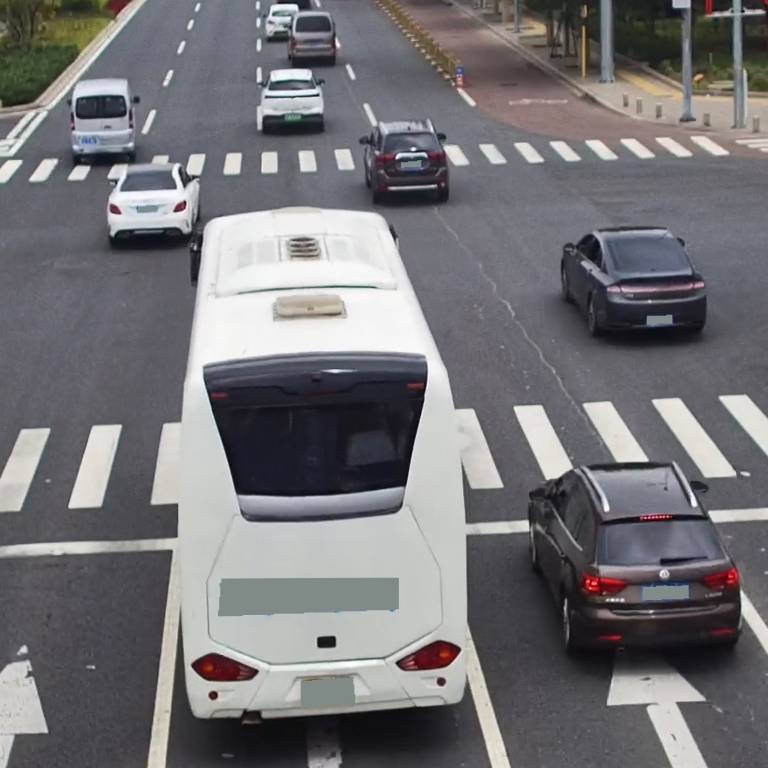} & 
			\includegraphics[width=\imgwidth]{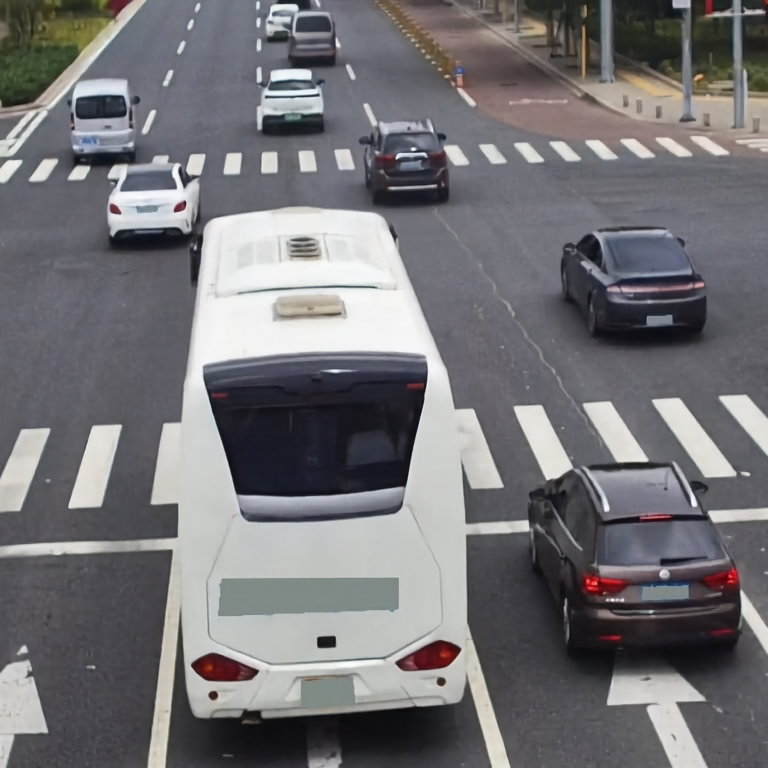} & 
			\includegraphics[width=\imgwidth]{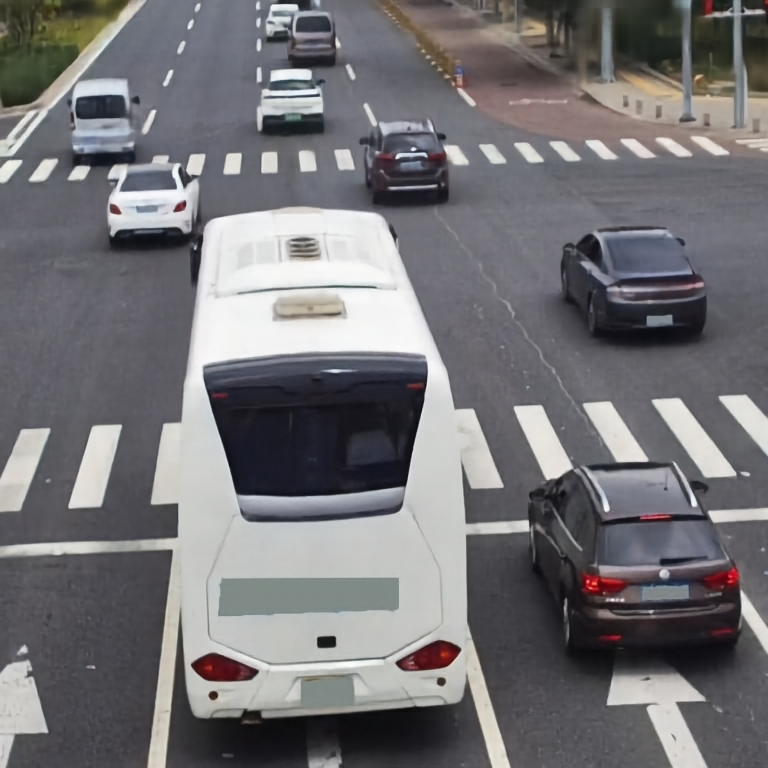} \\
			% --- 第4行：每张图底部的文字说明 ---
			\footnotesize DAIR-V2X & \footnotesize 0.27 / \textbf{0.93} & \footnotesize \textbf{0.26} / 2.63 & 
			\footnotesize DAIR-V2X & \footnotesize \textbf{0.38} / \textbf{1.02} & \footnotesize 0.44 / 11.59 &  
			\footnotesize DAIR-V2X & \footnotesize \textbf{0.11} / \textbf{0.23} & \footnotesize 0.15 / 0.89 \\
			
		\end{tabular}
		\vspace{0.1cm} % 调整上下子图间距
		\caption{Visualization results of the full SoM-MTM and its ablation version under different channel conditions. 
		%For the sake of clarity, all unmasked tokens (accounting for proportion $R_A$) are displayed with semi-transparency in their corresponding regions, while all lossy tokens (accounting for $R_L$) are shown without transparency. 
		The numbers below each image represent the distortion levels (MSE / $10^{-3}$) of image regions associated with successfully received and lossy tokens, respectively.}
		\label{pic_vis}
	\end{figure*}

	Through such more rigorous comparison, the superiority of the SoM-driven ERMoE method is further confirmed.
	When the number of activated experts is $1$, performance negatively correlates with $N_{tot}$, indicating that this sparse activation approach is unsuitable for this scenario.
	As $N_{act}$ increases, the computational cost rises significantly, yet its performance still struggles to surpass ours.
	The cost metrics quantitatively emphasize the efficiency advantage: the FLOPs of SoM-MTM lie between that of activating $1$ and $2$ experts, which is entirely acceptable. 
	Moreover, thanks to the external routing strategy enabling local model parallelism without cumbersome engineering optimization, its training time is actually the shortest, substantially reducing model development costs.
	
	\subsection{Hyperparameter Analysis}\label{SecExp-3}

	\subsubsection{Impact of Model Size (Scaling Analysis)}
	\begin{figure}[!t] % 强制图片位置
		\centering
		\includegraphics[width=0.9\linewidth]{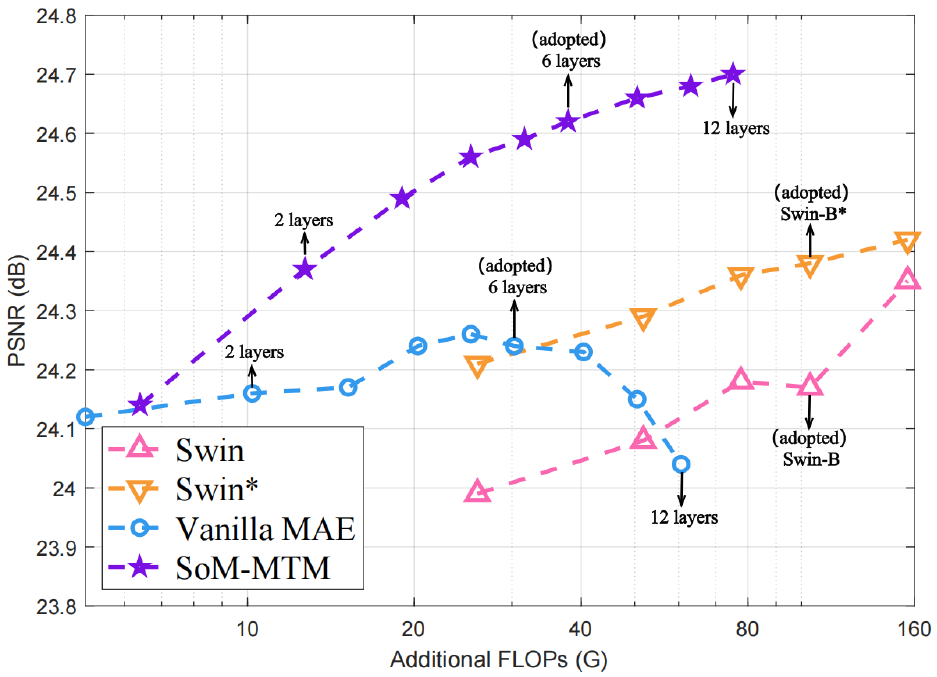} % 实际图片路径
		\caption{Model scaling performance at $R_A=30\%$.}
		\label{pic_hyper}
	\end{figure}
	As is well known, emerging AI architectures exemplified by Transformers exhibit strong scalability, with model capability positively correlated to its scale (scaling laws). Here, we similarly analyze the scaling effects of different schemes, with results shown in Fig. \ref{pic_hyper}. 
	What needs to be emphasized is that LCRN \& LCRN* employ a strictly fixed network structure thus lack scalability. 
	Swin \& Swin* 
	%represent baseline methods that organize packets by tokens or feature-channels, 
	rely on their native autoencoder architectures for contextual information learning and repairing. 
	They adopt a 4-layer progressive upsampling and downsampling structure and varying the number of stacked Transformers within the third layer according to \cite{SwinJSCC}.
	Vanilla MAE and SoM-MTM are both plug-in, utilizing variable-depth ViT and our custom-designed modules as the fundamental processing layers, respectively.
	Consistent with the previous setup, all schemes start from Swin-S, and the horizontal axis represents the additional FLOPs compared to it.

	It is obvious that SoM-MTM consistently realizes the optimal trade-off between model performance and computational cost.
	Baselines relying on the native recovery capability of autoencoders incur much higher costs than SoM-MTM due to differences in module arrangement, which is disadvantageous for CP that prioritize efficiency and end-to-end latency.
	Compared to others, Vanilla MAE does not exhibit the expected scaling curve. Its performance declines when the number of layers increases beyond a certain threshold ($6$ layers).  
	MAE is designed for pre-training encoders of conventional computer vision field, so that its transfer to feature-level CP system is not well-suited. 
	In contrast, SoM-MTM's performance trend fully aligns with scaling laws, demonstrating that our approach expands model capacity and elevates the upper bound of contextual learning. In practical applications, models of appropriate sizes can be selected based on computational resources and latency requirements (version with $6$ layers in our paper).
	
	\subsubsection{Impact of Attention Window Size}
	Meanwhile, for our Swin Transformer backbone, the window size for local attention mechanism is a critical factor. It determines the scope of context learning and requires the visual encoder to balance the compression of individual tokens with the semantic correlation among nearby tokens.
	
	Taking the situation with $R_A=30\%$ as an example, Table \ref{tab:win_size} presents the performance under different window sizes, reflecting our expected trade-off. When the sliding window for self-attention is too small, the vision field of is limited, hindering its ability to learn perceptual contextual information. Conversely, when the window becomes too large, it struggles to focus on relevant tokens, and tokens that are spatially distant instead become a source of interference. Ultimately, we select the window size of $4 \times 4$, which achieves the optimal result.
	\begin{table}[!t]  %window size实验表格
		\centering
		\caption{Model performance under different window sizes at $R_A=30\%$.}
		\label{tab:win_size}
		\renewcommand{\arraystretch}{1.35}
		\setlength{\tabcolsep}{6pt}
		\begin{tabular}{@{}c|cccc@{}}
			\specialrule{1.0pt}{0pt}{0pt} % 加粗线
			Window Size & $2^2$ & $4^2$ & $8^2$ & $16^2$ \\ \hline
			PSNR (dB) & 24.53&	\textbf{24.62} &	24.53&	23.80 \\
			\specialrule{1.0pt}{0pt}{0pt} % 加粗线
		\end{tabular}
	\end{table}

	\subsection{Supporting End-to-end Downstream Tasks}\label{SecExp-4}
	Previous results highlight that SoM-MTM excels at reconstructing raw perceptual information. Based on high-quality images, it can freely support various tasks in accordance with requirements. 
	While in certain situations, if a specific target task is predefined, the process can be simplified via another route: bypassing reconstruction of raw data and directly utilizing lightweight task heads to derive outputs from intermediate features. This approach reduces error propagation and exploits end-to-end gains, representing a common paradigm in CP.

\begin{figure}[!t] % 强制图片位置
		\centering
		\hspace*{\fill}
		\subfloat[Classification Top-1 Acc]{
			\includegraphics[width=0.46\linewidth]{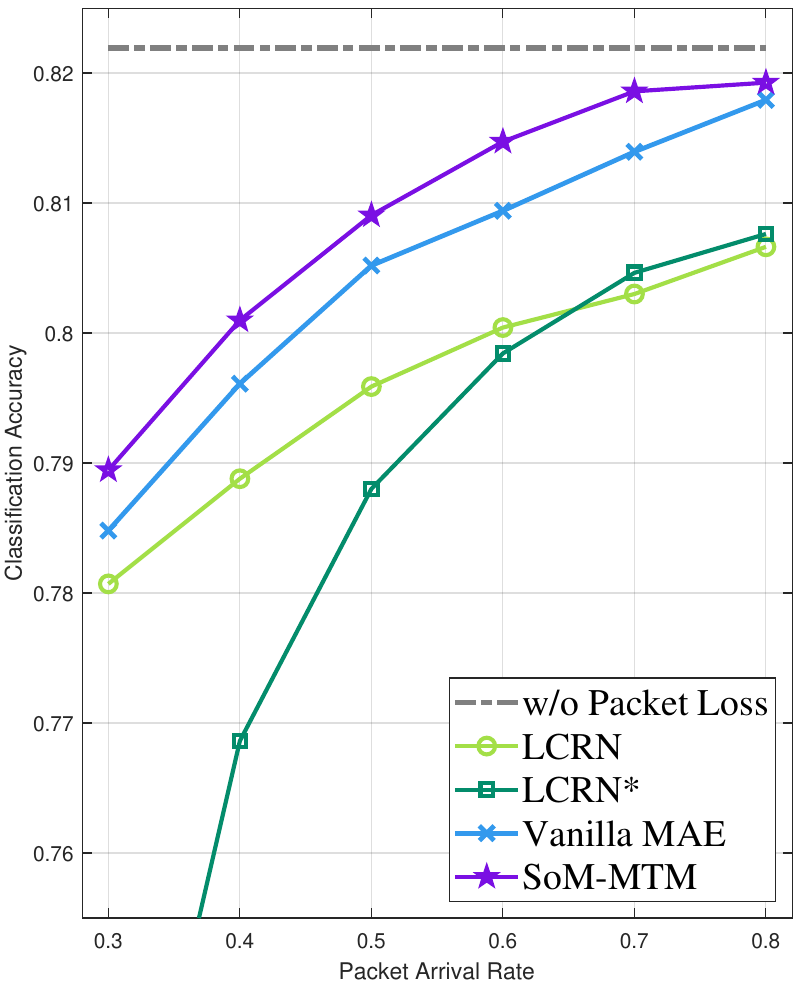} % 实际图片路径
			\label{pic11}
		}
		\subfloat[Segmentation mIOU]{
			\includegraphics[width=0.46\linewidth]{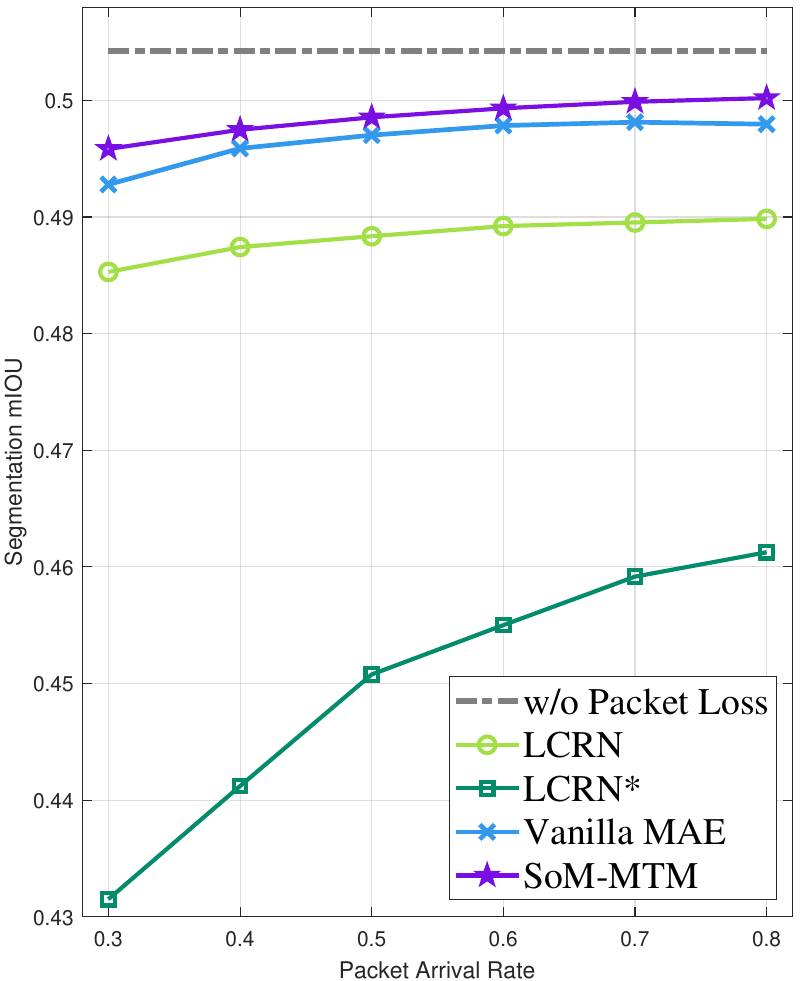} % 实际图片路径
			\label{pic22}
		}
		\hspace*{\fill}
		\caption{Comparison of end-to-end downstream task performance between the proposed SoM-MTM and baseline schemes.}
		\label{pic_perception}
	\end{figure}

	Fig. \ref{pic_perception} presents the outcome when directly handling downstream tasks. %based on the aforementioned technical route. 
	In this case, all schemes are pluggable, with the dataset and backbone selected detailed in Table \ref{tab:tasks}.
	Similarly, the proposed SoM-MTM achieves the best on both downstream tasks. For classification, its average Top-1 Accuracy across all tested points reaches 80.9\%; for semantic segmentation, the mIoU attains 49.9\%. Corresponding metrics for the second-ranked are 80.5\% and 49.7\%, respectively. 
	
	Additionally, we observe that as $R_A$ decreases (worse channel conditions), the decline in perceptual performance is very gradual. 
	Fig. \ref{pic_perception} also includes the theoretical upper bound, i.e., the situation without packet loss, where the corresponding metrics are 82.1\% and 50.4\%. The gap between our approach and this upper bound is minimal.
	It suggests that when only specific downstream tasks need to be accomplished without concern for raw perceptual data, intermediate feature information exhibits greater redundancy and stronger tolerance to packet-loss transmission.
	Under this circumstance, the designed working conditions of lossy channels are highly compatible. 
	Taking the example here, low-altitude drones or home service robots serve as the vision-node, simultaneously supporting multiple task-sides in dense access and dynamic channel conditions while performing classification or segmentation. 
	Even under highly adverse channel conditions with high packet loss rates, perceptual tasks can still be accomplished with high quality. 
	This greatly benefits multi-agent systems, meaning that future systems can support more devices collaborating under broader conditions through such connectivity without increasing the burden on communication networks.
	
	\section{Conclusion}\label{Sec-7}
	% 重复结论
	This paper has proposed SoM-MTM as an efficient plug-and-play model that can be flexibly and cost-effectively applied to task-node, significantly boosting cooperative perception performances. 
	Inspired by the MAE-style representation architecture, our model possesses the ability to learn perceptual contextual information. 
	By organizing data packets at the token level and employing suitable communication strategies, it is highly compatible with packet loss channels, substantially improving cooperation efficiency. 
	Furthermore,  we introduce  prior communication knowledge into SoM-MTM, which can strengthen the robustness of contextual features. Extensive experimental evaluations across diverse datasets and varying configuration conditions demonstrate its superiority. 
	Whether for generic image reconstruction or specialized downstream tasks, SoM-MTM achieves competitive perception performance while maintaining strong scalability.
	
	% Future Work
	Moreover, to further enhance collaboration between the vision-node and task-node and improve CP efficiency, we propose the following future research directions under the guidance of the SoM paradigm:
	i) Introduce an adaptive processing mechanism in the visual encoder to better coordinate with the task-node, enabling flexible operation under varying transmission conditions such as bandwidth, packet size, user count, and channel dynamics.
	ii) Extend the current token masking methodology to more complex tasks, such as integrating with multimodal perception and vision-language models (VLM), to explore the application potential of SoM-MTM for larger and more comprehensive AI models.
	
	\begin{comment}
		\section*{Acknowledgments}
		This should be a simple paragraph before the References to thank those individuals and institutions who have supported your work on this article.
	\end{comment}
	%\begingroup
	\fontsize{7.8pt}{8.1pt}\selectfont %修改REF部分大小
	
	\bibliographystyle{IEEEtran}
	\bibliography{myrefs.bib}
	
	%\endgroup
	
\end{document}